\pdfoutput=1
\documentclass[twocolumn,prd,a4paper,final,superscriptaddress,longbibliography]{revtex4-1}
\usepackage{amsmath,amssymb}
\usepackage{graphicx}
\usepackage{comment}
\usepackage{bm}
\usepackage[colorlinks=true,allcolors=blue]{hyperref}
\usepackage{color}
\usepackage{subfigure}
\usepackage[resetlabels]{multibib}
\newcites{SM}{SM bibliography}

\newcommand{\beq}{\begin{eqnarray}}
\newcommand{\eeq}{\end{eqnarray}}

\newcommand{\p}{\partial}
\newcommand{\Tr}{\qopname\relax o{Tr}}

\newcommand{\CP}{\mathbb{C}P}
\newcommand{\SU}{\mathrm{SU}}
\newcommand{\U}{\mathrm{U}}
\renewcommand{\d}{{\mathrm{d}}}
\renewcommand{\i}{\mathrm{i}}

\renewcommand{\Re}{\qopname\relax o{Re}}
\renewcommand{\Im}{\qopname\relax o{Im}}

\newcommand{\bS}{\mathbf{S}}
\newcommand{\bX}{\mathbf{X}}
\newcommand{\bQ}{\mathbf{Q}}
\newcommand{\bn}{\mathfrak{n}}
\newcommand{\bq}{\mathbf{q}}
\newcommand{\br}{\mathbf{r}}

\newcommand{\hS}{\hat{S}}
\newcommand{\hQ}{\hat{Q}}

\newcommand{\hdel}{\hat{\delta}}
\newcommand{\heta}{\hat{\eta}}
\newcommand{\hb}{\hat{b}}
\newcommand{\hn}{\hat{n}}
\newcommand{\hH}{\hat{H}}
\newcommand{\hT}{\hat{T}}

\newcommand{\calD}{{\cal D}}
\newcommand{\calS}{{\cal S}}

\newcommand{\calO}{{\cal O}}

\def\ket#1{\left|{#1}\right>}
\def\bra#1{\left<{#1}\right|}

\def\exv#1{\langle{#1}\rangle}

\begin{document}
\title{\texorpdfstring{$\CP^2$}{CP(2)} Skyrmion Crystals in an \texorpdfstring{$\SU(3)$}{SU(3)} Magnet with a Generalized Dzyaloshinskii-Moriya Interaction}
	
\date{April 5, 2022}
\author{Yuki Amari}
\affiliation{Department of Physics, Graduate School of Science, The University of Tokyo, Bunkyo, Tokyo 113-0033, Japan}
\affiliation{Bogoliubov Laboratory of Theoretical Physics, Joint Institute for Nuclear Research, Dubna 141980, Moscow Region, Russia}
\affiliation{Department of Physics, Faculty of Science and Technology, Tokyo University of Science, Noda, Chiba 278-8510, Japan}
\affiliation{Department of Mathematical Physics, Toyama Prefectural University, Kurokawa 5180, Imizu, Toyama, 939-0398, Japan}
\affiliation{Research and Education Center for Natural Sciences, Keio University, Hiyoshi 4-1-1, Yokohama, Kanagawa 223-8521, Japan}
\author{Yutaka Akagi}
\affiliation{Department of Physics, Graduate School of Science, The University of Tokyo, Bunkyo, Tokyo 113-0033, Japan}
\author{Sven Bjarke Gudnason}
\affiliation{Institute of Contemporary Mathematics, School of Mathematics and Statistics, Henan University, Kaifeng, Henan 475004, People’s Republic of China}
\author{Muneto Nitta}
\affiliation{Research and Education Center for Natural Sciences, Keio University, Hiyoshi 4-1-1, Yokohama, Kanagawa 223-8521, Japan}
\affiliation{Department of Physics, Keio University, Hiyoshi 4-1-1, Yokohama, Kanagawa 223-8521, Japan}
\author{Yakov Shnir}
\affiliation{Bogoliubov Laboratory of Theoretical Physics, Joint Institute for Nuclear Research, Dubna 141980, Moscow Region, Russia}
\affiliation{Institute of Physics, University of Oldenburg, D-26111 Oldenburg, Germany}

	\vspace{.5in}
	
	\begin{abstract}
		We study $\CP^2$ Skyrmion crystals in 
		the ferromagnetic $\SU(3)$ Heisenberg model with a generalization of the
		Dzyaloshinskii-Moriya interaction and the Zeeman term. 
		The model possesses two different types of Skyrmion crystals
		with unit-Skyrmions that can be interpreted as bound states of two
		half-Skyrmions or four quarter-Skyrmions. 
		Our study on $\CP^2$ Skyrmion crystals opens up the possibility for useful future applications since $\CP^2$ Skyrmions have more degrees of freedom than the usual $\CP^1$ (magnetic) Skyrmions.
	\end{abstract} 
	\maketitle

	\section{Introduction}\label{sec:intro}
	
	Skyrmions in their original incarnation were invented by Skyrme as a
	simple topological model of nuclei \cite{Skyrme:1962vh}, but were first taken
	more seriously after Witten showed that they are the baryons of
	large-$N_c$ quantum chromodynamics (QCD) \cite{Witten:1983tx} and have
	led to many qualitative results \cite{MantonsBook,multifaceted}.
	Insights for the interactions of Skyrmions at large separations were
	subsequently found in a two-dimensional toy model, called the
	baby-Skyrme model \cite{Bogolubskaya:1989ha,doi:10.1007/BF01045888,Piette:1994ug}.
	Recently, similar topological structures known as magnetic
	Skyrmions \cite{Bogdanov:1989,Bogdanov:1995} have received quite intensive focus due to their
	realizations in the laboratory in 
	chiral~\cite{doi:10.1126/science.1166767,doi:10.1038/nature09124,doi:10.1038/nphys2045}
	or noncentrosymmetric~\cite{Kurumaji2019,Hirschberger2019,Khanh2020,Yasui2020}
	magnets and their possible applications 
	as components for data storage with low energy consumption
	\cite{doi:10.1038/nnano.2013.29} 
	(see Ref.~\cite{Nagaosa2013} for a review).

	Skyrmions in two-dimensional materials are topological solitons with
	the target of a 2-sphere, which is parametrized by a magnetization
	vector of fixed length. The topological charge or degree comes from
	considering only finite energy configurations, which forces the
	magnetization vector to be a constant at asymptotic distances and
	therefore the topology is that of maps between two spheres:
	$\pi_2(S^2)=\mathbb{Z}\ni N$. 
	Nontrivial topology, however, does not ensure 
	that Skyrmions are actually
	realizable in a material. It is also necessary that there is some
	stabilizing mechanism at work.
	Skyrmions in chiral magnets are stabilized by the
	Dzyaloshinskii-Moriya (DM) interaction term --- stemming from
	spin-orbit coupling (SOC) --- which stabilizes Skyrmions with one 
	chirality, but not the other. 
	Skyrmions in chiral magnets have a fixed vector chirality
	\cite{doi:10.1126/science.1166767}, whereas this is a degree of
	freedom in noncentrosymmetric materials and the stabilizing
	mechanism at work is also different. 
	In the latter materials, the stabilization of Skyrmions is due to
	frustration, magnetic anisotropy~\cite{PhysRevLett.108.017206, Leonov2015,Amoroso2020},
	and multiple-spin interactions mediated by itinerant electrons~\cite{Akagi2012,Ozawa2017,Hayami2017a}, instead of the DM term. 
	
	A two-dimensional sphere ($S^2$) can also be viewed as a complex
	projective plane $(\CP^1)$. Although planar Skyrmions cannot be
	topologically stable with higher-dimensional spheres $S^n$, $n>2$, for
	their target space, they can be topological for $\CP^n$ with $n\geq1$.
	The $\CP^n$ model was proposed about half a century ago~\cite{Eichenherr:1978qa,Golo:1978de,Cremmer:1978bh} 
	and has been studied in quantum field theory, 
	as the (1+1)-dimensional $\CP^n$ model shares 
	various properties with (3+1)-dimensional gauge theories, such as a dynamical mass gap, asymptotic freedom, and 
	instantons (spacetime analogs of planar Skyrmions)~\cite{DAdda:1978vbw,Witten:1978bc}.
	In condensed matter physics, 
	the $\CP^n$ model 
	has been studied for a new quantum phase transition called 
	deconfined criticality~\cite{Senthil:2003eed,Nogueira:2013oza},
	and proposed to be realized in 
	ultracold atomic gases~\cite{Laflamme:2015wma}, 
	multiband superconductors \cite{garaud2011topological,Garaud:2012pn,benfenati2022},
	and $S=1$ spin systems~\cite{Papanicolaou1988, Batista2004, Tsunetsugu2006, Lauchli2006, Toth2010, penc2011spin, Bauer2012} where solitons (Skyrmions)~\cite{Ivanov2003, Ivanov2007, ivanov2008pairing, Galkina2015, Ueda_2016} and vortices~\cite{Ivanov2003, Takano2011, Grover2011, Xu2012, Hu2014, Remund2022} can emerge.
	A physically relevant and interesting minimal extension is the case of
	the $\CP^2$ target space, which appears as the order parameter space of 
	an effective model of the 
	spin-1 Bose-Hubbard model~\cite{imambekov2003spin,ivanov2008pairing}.
	In addition, an $\SU(3)$ SOC can be induced by applying a laser beam to ultracold atomic gases~\cite{PhysRevA.81.053403,Dalibard:2010ph,Goldman:2013xka,Zhai:2014gna,Note1}.
	\footnotetext[1]{Stabilization of three-dimensional Skyrmions in ultracold atomic gases with an SOC was proposed in Ref.~\cite{Kawakami:2012zw}.}
	Therefore, ultracold atom systems offer a promising candidate to realize $\CP^2$ Skyrmions.
	
	Skyrmions realized in nature are often in the form of crystals.
	Skyrmion crystals were first considered in the three-dimensional Skyrme model~\cite{Klebanov:1985qi,Kugler:1989uc} and at finite (large) density, there is a transition from unit Skyrmions to half-Skyrmions~\cite{parkvento}.
	Arrays or crystals of magnetic Skyrmions for normal $\CP^1$ Skyrmions are also 
	realized in 
	magnets~\cite{doi:10.1126/science.1166767,doi:10.1038/nature09124,doi:10.1038/nphys2045, Kurumaji2019,Hirschberger2019,Khanh2020,Yasui2020}.
	On the other hand, such magnetic Skyrmion crystals have been considered in classical spin systems or $\mathbb{C}P^1$ magnets.
	Since Skyrmions with $\CP^2$ target space are also topological, but have more internal structure, the realization of a $\CP^2$ Skyrmion crystal in some material may have useful future applications.
	
	The purpose of this Letter is to propose the possibility of 
	Skyrmion crystals with the degrees of freedom of $\CP^2$.
	We find that
	there are two different crystal types, depending on a single free
	parameter.
	The parameter
	has to be small enough (below a critical value) for the Skyrmions to
	exist and not be energetically disfavorable to the ferromagnetic
	phase.

	\section{Model}
	
	We consider a low-energy effective Hamiltonian of the spin-1
	Bose-Hubbard model with an $\SU(3)$ SOC on a square lattice. 
	Let $\hat{S}^a_i (a=x,y,z)$ and $\hat{T}^\alpha_i (\alpha=1,2,...,8)$ be the spin-1 and $\SU(3)$ spin operators defined on site $i$, respectively. In terms of the operators, the Hamiltonian
	is given by
	\begingroup
	\allowdisplaybreaks
	\begin{align}
		\hat{H} &= \hat{H}_{\SU(3)} + \hat{H}_{\rm DM} + \hat{H}_{\rm Zeeman},\label{eq:H}\\
		\hat{H}_{\SU(3)} &= \frac{J}{2}\sum_{\langle i,j\rangle}\sum_{\alpha=1}^8
		\hat{T}_i^\alpha \hat{T}_j^\alpha,\\
		\hat{H}_{\rm DM} &= J\sum_{\langle i,j\rangle}\sum_{\alpha,\beta,\gamma=1}^8
		f_{\alpha\beta\gamma} A_{i,j}^\alpha \hat{T}_i^\beta \hat{T}_j^\gamma,\\
		\hat{H}_{\rm Zeeman} &= -h\sum_i\hS^z_i,\label{eq:HZeeman}
	\end{align}
	\endgroup
	where $\hat{H}_{\SU(3)}$, $\hat{H}_{\rm DM}$, and $\hat{H}_{\rm Zeeman}$ are the ferromagnetic $\SU(3)$ Heisenberg term ($J<0$), the 
	generalized DM interaction term~\cite{Akagi:2021dpk}, and the Zeeman interaction, respectively. 
	Here, the sum $\langle i,j\rangle$ is taken over the nearest-neighbor sites, 
	$f_{\alpha\beta\gamma}$ are the structure constants of $\SU(3)$ 
	defined as
	$f_{\alpha\beta\gamma}=-\frac{\i}{4}\Tr(\lambda_\alpha[\lambda_\beta,\lambda_\gamma])$
	where $\lambda_\alpha$ $(\alpha=1,2,\ldots,8)$ are the Gell-Mann matrices,
	and $A_{i,j}^\alpha=\frac12\Tr(\lambda_\alpha A_{i,j})$ is the $\SU(3)$
	gauge potential.
	The $\SU(3)$ spin operator can be written as a product of the spin-1 operators, and inversely the spin-1 operators can be defined by the $\SU(3)$ spin operators. In this Letter, we use the gauge defining the spin-1 operators as
		\beq
		\hat{\bS}_i =
		\left(\frac{\hat{T}_i^1+\hat{T}_i^6}{\sqrt{2}},
		\frac{\hat{T}_i^2+\hat{T}_i^7}{\sqrt{2}},
		\frac{\hat{T}_i^3+\sqrt{3}\hat{T}_i^8}{2}\right).
		\eeq
	To study Skyrmion crystals in this model, we employ the variational approach with an $\SU(3)$ coherent state,
\beq
\ket{\mathbf{Z}}=\otimes_i\ket{\mathbf{Z}_i},~ \ket{\mathbf{Z}_i}= Z^m_i\ket{m}_i \:\;\; {\rm such that} \;\;\: \mathbf{Z}^\dagger_i \mathbf{Z}_i =1.
\eeq
Here, $\ket{m}_i \equiv \ket{S=1,m}_i$, ($m=0$, $\pm 1$) are 
the eigenstates of $\hS^z_i$, and $\mathbf{Z}_i=(Z^1_i, Z^0_i, Z^{-1}_i)^T$. The state represents an arbitrary spin-1 state.  The classical Hamiltonian to be minimized is given by the expectation value of the quantum Hamiltonian \eqref{eq:H} in the coherent state
\begin{align}
	H&=\bra{\mathbf{Z}}\hat{H}\ket{\mathbf{Z}} \notag\\
	&=\frac{J}{2}\sum_{\langle i,j\rangle}\left[\sum_{\alpha=1}^8
	n_i^\alpha n_j^\alpha
	+\sum_{\alpha,\beta,\gamma=1}^8
	2f_{\alpha\beta\gamma} A_{i,j}^\alpha n_i^\beta n_j^\gamma\right]\notag\\
	&\quad-\frac{h}{2}\sum_i(n_i^3 + \sqrt{3}n_i^8),
	\label{classical_Hamiltonian}
\end{align} 
	where the field $n_i^\alpha$ is the expectation value of the $\SU(3)$ spin operator on
	the lattice site $i$ 
	defined as
	\beq
	n_i^\alpha = \mathbf{Z}_i^\dag\lambda_\alpha \mathbf{Z}_i.
	\eeq
	The field $n_i^\alpha$ satisfies
	\beq
	\sum_{\alpha=1}^8n^\alpha_i n^\alpha_i = \frac43,\quad
	\sum_{\alpha,\beta=1}^8d_{\alpha\beta\gamma}n^\alpha_i n^\beta_i
	= \frac32n^\gamma_i,
	\label{eq:constraint}
	\eeq
	where $d_{\alpha\beta\gamma}=\frac14\Tr(\lambda_\alpha,\{\lambda_\beta,\lambda_\gamma\})$
	are the symmetric symbols of $\SU(3)$.

	For the $\SU(3)$ gauge potential $A_{i,j}$, we use the form 
	\beq
	A_{i,i\pm\hat{x}} = \pm\frac{\kappa}{J}(\lambda_1+\lambda_6),\quad
	A_{i,i\pm\hat{y}} = \pm\frac{\kappa}{J}(\lambda_2+\lambda_7),
	\label{eq:gaugepot}
	\eeq
	where $\hat{x}$ and $\hat{y}$ denote the bond vectors of the length of the lattice spacing, and $\kappa$ is a constant.
	The $\SU(3)$ SOC with Eq.~\eqref{eq:gaugepot} can be recognized as a generalization of the Rashba SOC, which can be engineered in cold atom systems~\cite{PhysRevA.81.053403,Note2}\footnotetext[2]{It may still be an experimental challenge to control the DM term in two (spatial) dimensions.}.
	
	Skyrmions are topological solitons with an integer topological charge
	that is the degree of the map from two-dimensional space $\mathbb{R}^2$
	to the $\CP^2$ target space: $\pi_2(\CP^2)=\mathbb{Z}\ni N$.
	The topological charge $N$ can be computed by integrating the topological
	charge density
	\beq
	N = 
	\frac{\i}{32\pi}
	\int\d^2x\;\epsilon_{jk}\Tr(\bn[\p_j\bn,\p_k\bn]),
	\eeq
	with the color field $\bn=\sum_{\alpha=1}^8n^\alpha\lambda_\alpha$, for a continuous
	limit of the field $n^\alpha_i$.
	In the lattice model, the topological charge density is analogously
	given by $N = \sum_i N_i$ with
	\begin{equation}
		N_i = -\frac{1}{16\pi}\sum_{\alpha,\beta,\gamma=1}^8f_{\alpha\beta\gamma}
		n_i^\alpha(n_{i+\hat{x}}^\beta - n_{i-\hat{x}}^\beta)
		(n_{i+\hat{y}}^\gamma - n_{i-\hat{y}}^\gamma).
	\end{equation}
	The topological charge density $N_i$ can also be interpreted as the
	$\SU(3)$ scalar spin chirality. 

	In this Letter, we set $J=-1$ without loss of generality. In addition, for studying long-wavelength excitations such as Skyrmions, it is reasonable to fix one more parameter, because if we take the continuum limit, we can scale away two coupling constants. Therefore, we here fix $\kappa=0.2$ and let $h$ be the free parameter.

	\subsection{Numerical method}
	
	Our numerical method is described as follows.
	We first perform the minimization of the Hamiltonian \eqref{classical_Hamiltonian} with periodic boundary conditions using an unbiased, single-update simulated annealing method
	with randomly generated initial configurations, increasing the inverse
	temperature $\beta$ by a factor of $1.01$ at every $10^4$--$10^5$
	Monte Carlo steps, until $\beta$ reaches
	$\beta^{\rm max}\sim 10^2$--$10^4$.
	After that, we use the nonlinear conjugate gradients method to solve
	the equations of motion
	\begin{equation}
		\begin{split}
			\frac{\p n_i^\alpha}{\p\Re Z_i^m}\frac{\p H}{\p n_i^\alpha}
			- \omega_i\Re Z_i^m = 0,\\
			\frac{\p n_i^\alpha}{\p\Im Z_i^m}\frac{\p H}{\p n_i^\alpha}
			- \omega_i\Im Z_i^m = 0,
		\end{split}
	\end{equation}
	to obtain configurations with precise energies.
	The parameters $\omega_i$ are Lagrange multipliers at each lattice
	site $i$.
	
	\section{\texorpdfstring{$\CP^2$}{CP(2)} Skyrmion crystals} 

\begin{figure}[!htp]
	\centering
	\includegraphics[width=\linewidth]{{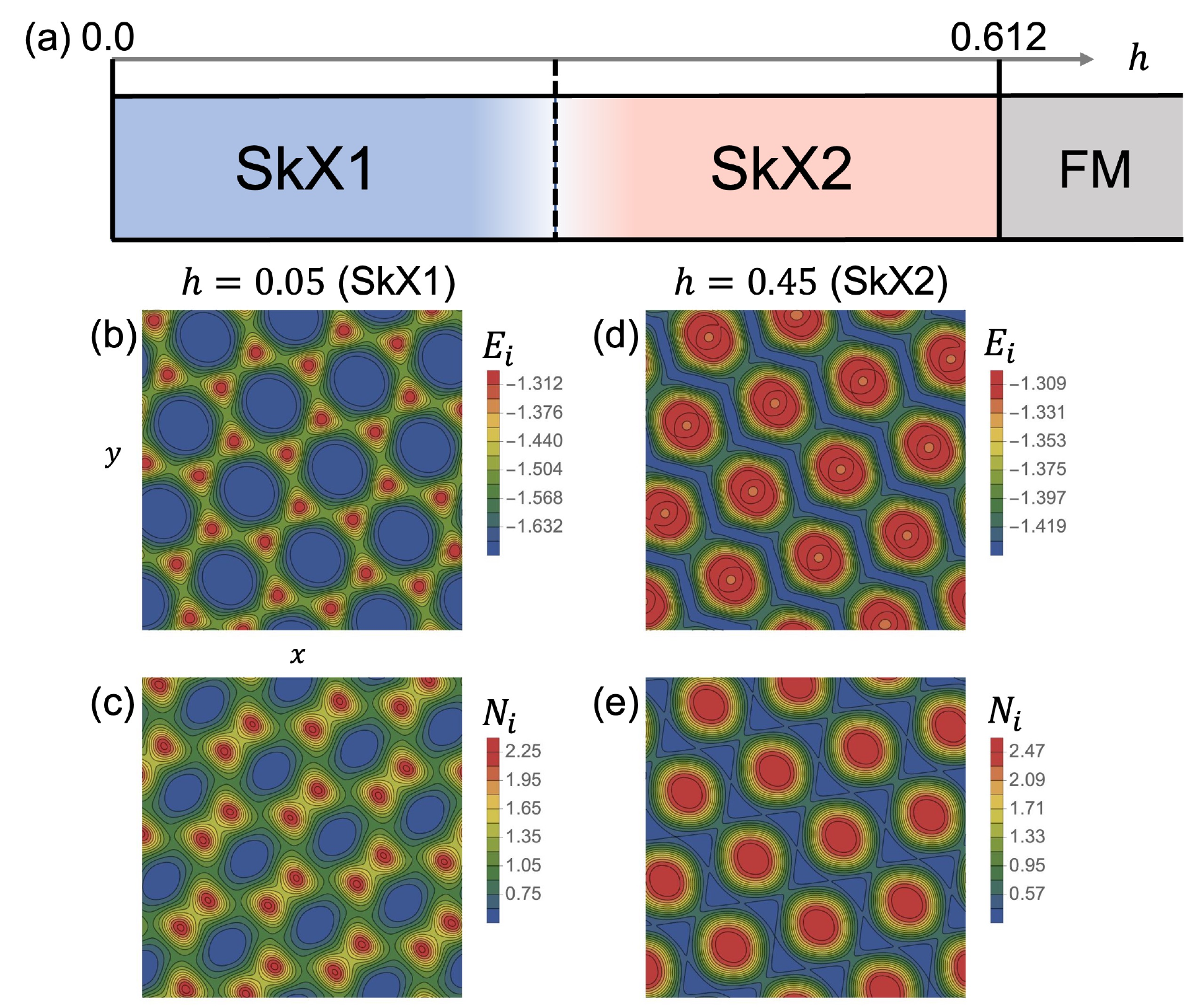}}
	\caption{Phase diagram showing typical Skyrmion crystal
		configurations of SkX1 and SkX2 types. (b) and (d)
		show the energy density of SkX1 and SkX2, respectively,
		whereas (c) and (e) show the topological charge
		distribution.
		The simulations depicted here are made using $32^2$ lattice sites.}
	\label{fig1}
\end{figure}

We are now ready to explore Skyrmion crystals in the
model \eqref{classical_Hamiltonian}, which after 
fixing the parameters only possesses
one free parameter, i.e.,~$h$. Varying $h$, we find three types of
configurations: a Skyrmion crystal of the first kind (SkX1), a Skyrmion crystal of the second kind (SkX2),
and the ferromagnetic state (FM) (see Fig.~\ref{fig1}).
We can see from the figure that SkX1 on a square lattice
is almost a honeycomb lattice of half-Skyrmions, whereas SkX2
consists of a triangular lattice of Skyrmions with unit
topological charge.

For convenience, we have subtracted off a constant from the energy so
as to render the energy of the FM phase equal to zero.
More precisely, we compute the energy as
\beq
E=H-\sum_i\left(\frac{4}{3}J - h\right),
\eeq
with the Hamiltonian $H$. 
We determine the critical value of $h$ for the phase transition to the
FM phase by calculating the Skyrmion energy. The critical value
$h_{\rm crit}\approx0.612$ is the value of the Zeeman coupling for
which the energy of the Skyrmion becomes positive. 

\begin{figure}[!htp]
	\centering
	\includegraphics[width=\linewidth]{{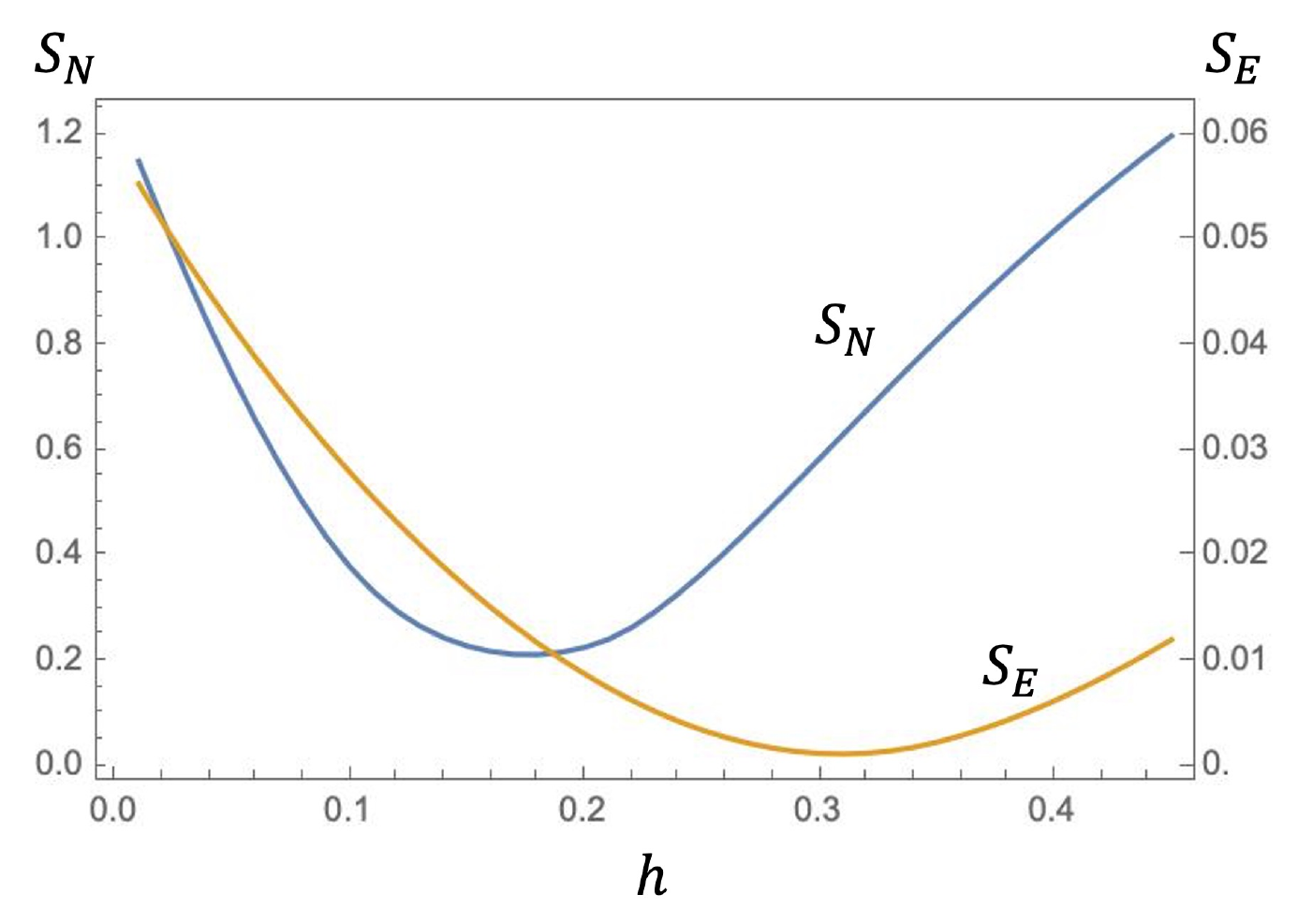}}
	\caption{Configurational entropy of the energy density $S_E$ and topological
		charge density $S_N$.}
	\label{fig2}
\end{figure}
On the other hand, SkX1 and SkX2 are connected via a crossover.
In order to determine how they are distinguished, 
we propose to compute the configurational entropy (CE) of the energy and
the topological charge densities, respectively~\cite{Gleiser:2011di,Gleiser:2015aav,bazeia2019configurational,bazeia2021configurational}:
\begin{equation}
	\begin{split}
		&S_{E(N)} = -\sum_lf_{E(N)}(\bq_l)\log f_{E(N)}(\bq_l),\\
		&f_{E(N)}(\bq_l) = \frac{|F_{E(N)}(\bq_l)|^2}{\sum_i|F_{E(N)}(\bq_i)|^2}.
	\end{split}
\end{equation}
Here, $F_{E}$ and $F_{N}$ are the discrete Fourier transform of the energy and
topological charge densities, respectively,
i.e.,~$F_{E(N)}(\bq_l)=\mathcal{N}^{-1}\sum_jE_j(N_j)e^{\i\bq_l\cdot\br_j}$, where
$\bq_l$ is the momentum conjugate to $\br_l$ and $\cal N$ is the number of lattice sites.

We find that the minimum of the CE signals a
change in the unit Skyrmion's local structure. In particular, starting
at small values of $h$, both the energy and topological charge
densities have two peaks close to their respective minima. These two peaks break up
into four peaks, which we may interpret as half-Skyrmions becoming
quarter-Skyrmions in SkX2. 
This transition happens for smaller values of $h$ for
the topological charge density (viz.,~$h_{\rm crit}^{\rm SkX1/2}\approx0.18$)
as compared to the same happening for the energy density
(viz.,~$h_{\rm crit}^{\rm SkX1/2}\approx0.31$).
Hence, there is not a clear-cut critical value of $h$ for the
transition between SkX1 and SkX2, but rather a range with a
kind of crossover between the two 
$\CP^2$ Skyrmion crystals.

\begin{figure*}[!htp]
	\centering
	\includegraphics[width=\linewidth]{{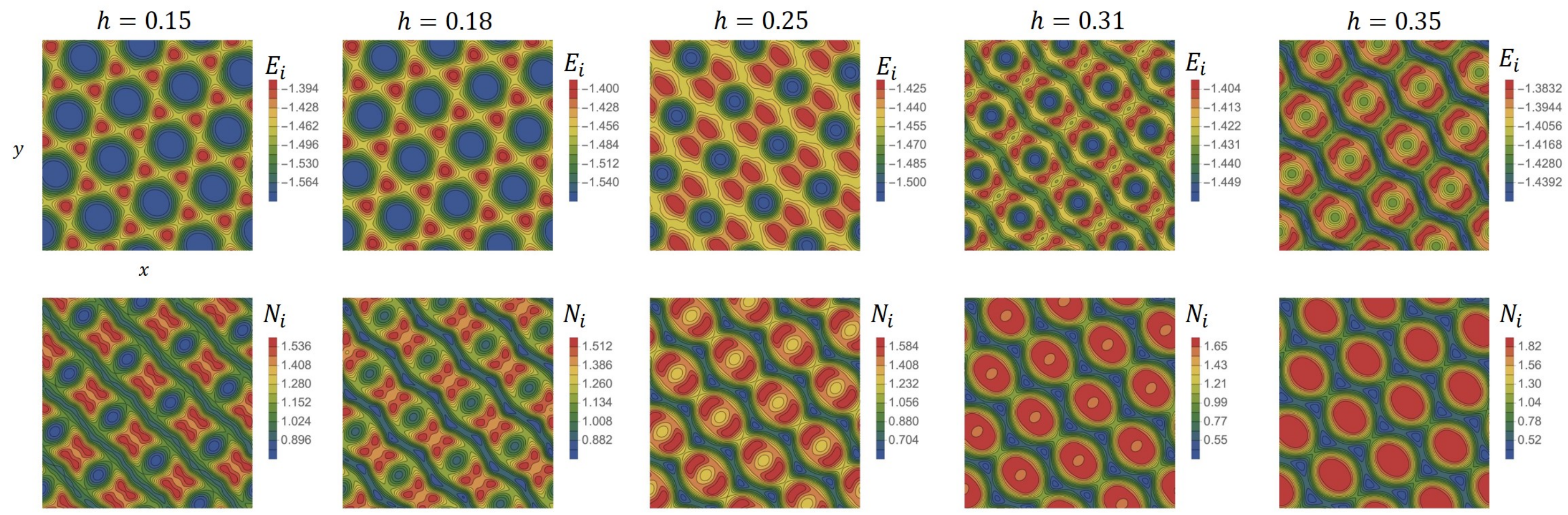}}
	\caption{The energy and topological charge densities of conﬁgurations
		for $h=0.15, 0.18, 0.25, 0.31$, and $0.35$. The upper panels
		represent the energy density and the lower ones show the
		topological charge density.} 
	\label{fig3}
\end{figure*}
\begin{figure*}[!htp]
	\centering
	\includegraphics[width=\linewidth]{{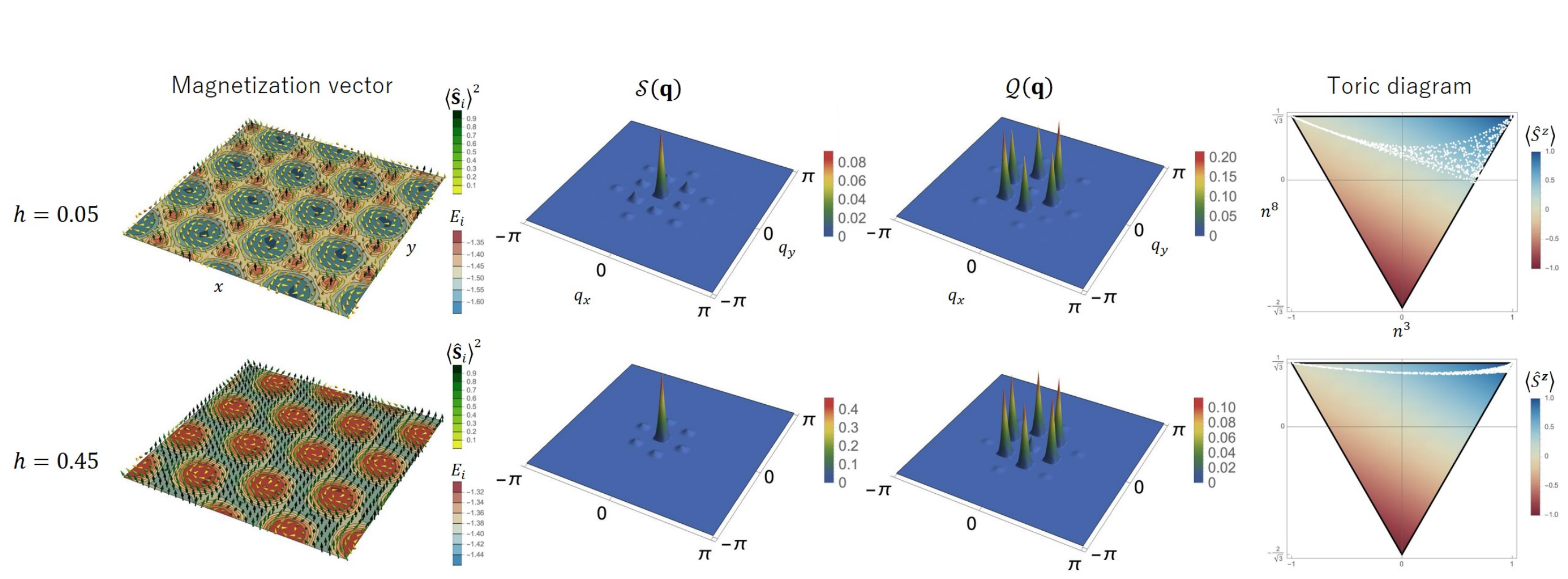}}
	\caption{Magnetization vector, 
		dipole structure factor $\mathcal{S}(\bq)$, quadrupole
		structure factor $\mathcal{Q}(\bq)$, and configurational 2-cycle of the Skyrmions
		shown on the toric diagram of $\CP^2$. The upper panels show the
		case of SkX1 
		with $h$ chosen as $h=0.05$, and the lower
		panels show SkX2 with $h=0.45$.
		The color of the magnetic vectors illustrates their length, whereas
		the background color corresponds to the energy density.
		The color of the toric diagram represents instead the value of the
		third spin-1 component, i.e.,~$\langle \hS^z\rangle=(n^3+\sqrt{3}n^8)/2$.
	}
	\label{fig4}
\end{figure*}

We will now discuss some further properties of the Skyrmion crystals
 (see Fig.~\ref{fig3} for the evolution of the two types of Skyrmion
crystals as functions of $h$, where the energy and topological charge
densities are shown as the top and bottom rows of the figure).
For the final properties, we define 
\beq
\mathcal{X}(\bq) = \mathcal{N}^{-2}\sum_{j,k}\langle\hat{\bX}_j\rangle\cdot
\langle\hat{\bX}_k\rangle e^{\i\bq\cdot(\br_j-\br_k)},
\eeq
where $\mathcal{X}=\mathcal{S}$ (with $\hat{\bX}=\hat{\bS}$) is the 
dipole structure
factor and $\mathcal{X}=\mathcal{Q}$ (with $\hat{\bX}=\hat{\bQ}$) is the
quadrupole structure factor. 
Here, the expectation value of the quadrupole operator $\hat{\bQ}_i$ is given by
\beq
\langle\hat{\bQ}_i\rangle =
\begin{pmatrix}
	\exv{\hat{Q}_i^{x^2-y^2}}\\
	\exv{\hat{Q}_i^{r^2-3z^2}}\\
	\exv{\hat{Q}_i^{xy}}\\
	\exv{\hat{Q}_i^{xz}}\\
	\exv{\hat{Q}_i^{yz}}
\end{pmatrix} =
\begin{pmatrix}
	n_i^4\\
	\frac{\sqrt{3}n_i^3-n_i^8}{2}\\
	n_i^5\\
	\frac{n_i^1-n_i^6}{\sqrt{2}}\\
	\frac{n_i^2-n_i^7}{\sqrt{2}}
\end{pmatrix},
\eeq
where $\hQ^{ab}_i=\hS^a_i\hS^b_i+\hS^b_i\hS^a_i-4\delta_{ab}/3$~\cite{penc2011spin}.
The middle of Fig.~\ref{fig4} shows the dipole structure factor $\mathcal{S}(\bq)$ and quadrupole structure factor $\mathcal{Q}(\bq)$.
As usually observed in $\CP^1$ Skyrmion crystals, $\mathcal{S}(\bq)$ shows the sharp peak at the $\Gamma$-point and small triple-$q$ structures with $|\bq| \equiv q_S$ under a magnetic field.
On the other hand, as the characteristic feature of $\CP^2$ Skyrmion crystals, sharp triple-$q$ structures with $|\bq| = q_S$ appear in $\mathcal{Q}(\bq)$.
One can also see that SkX1 has a higher triple-$q$ structure with $|\bq| >q_S$ in $\mathcal{S}(\bq)$. 
Another difference between SkX1 and SkX2 is the position of higher triple-$q$ peaks with $|\bq| >q_S$ in $\mathcal{Q}(\bq)$.

Figure~\ref{fig4} also shows the magnetization vector 
as well as the toric $\CP^2$
diagram with the 2-cycle of the Skyrmions charted out as white
points.
Since the 2-cycle is not concentrated on
a single straight line on the toric
$\CP^2$ diagram, we can see that the Skyrmions are genuine $\CP^2$
Skyrmions and not simply a $\CP^1$ Skyrmion embedded into $\CP^2$.
In SkX1, 
the magnetization vectors are longest at the core of
fractional Skyrmions and point in the $\hat{z}$ direction, whereas
they have vanishing length in SkX2.

\section{Conclusion and discussion}

In this Letter, we have proposed the possibility of $\CP^2$ Skyrmion
crystals and have found two different 
types of crystals in 
an $\SU(3)$ spin system with Zeeman and the generalized DM terms as the stabilizing
agent.
To obtain additional information on the unit Skyrmion's local structure,
we have computed the configurational entropy of the energy and topological charge densities.
Their minima correspond to bound states of two
half-Skyrmions and four quarter-Skyrmions.
As a characteristic feature of $\CP^2$ Skyrmion crystals,
we have found a triple-$q$ structure in the quadrupole structure factors.
Since $\CP^2$ Skyrmions have more internal structures than $\CP^1$ Skyrmions,
it is of great interest to explore emergent phenomena in $\CP^2$ Skyrmion crystals 
and consider their applications in future nanotechnology. 
In addition, we expect that $\CP^2$ Skyrmion crystals are relevant even in high-energy physics, e.g., dense quark matter possessing $\SU(3)$ ferromagnetism~\cite{kobayashi2014color}.

While we have studied Skyrmion crystals in a low-energy effective model of the simplest spin-1 Bose-Hubbard model with artificial gauge potentials and the Zeeman interaction, we expect that the Skyrmion crystals survive even if we slightly introduce the spin-dependent interaction in the spin-1 Bose-Hubbard model that appears in the system of spinor BECs~\cite{stamper2013spinor,Kawaguchi:2012ii}. To show this, is the most important future problem. 
While we have considered the linear Zeeman term, the quadratic Zeeman term splits even a singly isolated Skyrmion into fractional Skyrmions (merons)
\cite{Akagi:2021lva}, 
and thus investigating a Skyrmion crystal with the quadratic Zeeman term would be an interesting future direction of research.
It is also of great interest to consider Skyrmion crystals appearing in a pseudo-spin system, i.e., a mixture of three species of bosons~\cite{Gra__2014}.
Other interesting directions are 
to investigate the properties at finite temperature~\cite{Stoudenmire2009, Yamamoto2020, Tanaka2020, Remund2022} and
to study Skyrmion crystals in the anti-ferromagnetic $\SU(3)$ Heisenberg model with a generalized DM interaction, of which the continuum counterpart is a nonlinear sigma model on the flag manifold $\SU(3)/\U(1)^2$~\cite{Ueda_2016, Amari:2017qnb, Amari:2018gbq,Affleck:2021jls,Takahashi:2021lvg}.

\subsection*{Note added}

Recently, we became aware of Ref.~\cite{Zhang:2022lyz}, which has some overlap with our results, in particular, the
proposal to realize $\CP^2$ Skyrmion crystals, albeit in different physical
systems. In particular, 
while Ref.~\cite{Zhang:2022lyz} utilizes
frustration to stabilize the Skyrmions, we 
considered the generalized DM interaction on a square lattice. 

\begin{acknowledgments}
	The authors would like to thank N.~Sawado for
	useful discussions. 
	The work of Y.~Akagi is supported by JSPS KAKENHI Grant No.~JP20K14411
	and JSPS Grant-in-Aid for Scientific Research on Innovative Areas
	``Quantum Liquid Crystals'' (KAKENHI Grants No.~JP20H05154 and No.~JP22H04469).
    S.~B.~G.~thanks the Outstanding Talent Program of Henan University and
the Ministry of Education of Henan Province for partial support.
	The work of S.~B.~G.~is supported by the National Natural Science
	Foundation of China (Grants No.~11675223 and No.~12071111).
	The work of M.N.~is supported in part by JSPS Grant-in-Aid for
	Scientific Research (KAKENHI Grants No.~JP18H01217 and No.~JP22H01221).
	The computations in this
	paper were run on the ``GOVORUN" cluster supported by the LIT, JINR.
\end{acknowledgments}

\bibliographystyle{apsrev4-1}
\bibliography{refs.bib}

\begin{thebibliography}{4}%
\makeatletter
\providecommand \@ifxundefined [1]{%
 \@ifx{#1\undefined}
}%
\providecommand \@ifnum [1]{%
 \ifnum #1\expandafter \@firstoftwo
 \else \expandafter \@secondoftwo
 \fi
}%
\providecommand \@ifx [1]{%
 \ifx #1\expandafter \@firstoftwo
 \else \expandafter \@secondoftwo
 \fi
}%
\providecommand \natexlab [1]{#1}%
\providecommand \enquote  [1]{``#1''}%
\providecommand \bibnamefont  [1]{#1}%
\providecommand \bibfnamefont [1]{#1}%
\providecommand \citenamefont [1]{#1}%
\providecommand \href@noop [0]{\@secondoftwo}%
\providecommand \href [0]{\begingroup \@sanitize@url \@href}%
\providecommand \@href[1]{\@@startlink{#1}\@@href}%
\providecommand \@@href[1]{\endgroup#1\@@endlink}%
\providecommand \@sanitize@url [0]{\catcode `\\12\catcode `\$12\catcode
  `\&12\catcode `\#12\catcode `\^12\catcode `\_12\catcode `\%12\relax}%
\providecommand \@@startlink[1]{}%
\providecommand \@@endlink[0]{}%
\providecommand \url  [0]{\begingroup\@sanitize@url \@url }%
\providecommand \@url [1]{\endgroup\@href {#1}{\urlprefix }}%
\providecommand \urlprefix  [0]{URL }%
\providecommand \Eprint [0]{\href }%
\providecommand \doibase [0]{http://dx.doi.org/}%
\providecommand \selectlanguage [0]{\@gobble}%
\providecommand \bibinfo  [0]{\@secondoftwo}%
\providecommand \bibfield  [0]{\@secondoftwo}%
\providecommand \translation [1]{[#1]}%
\providecommand \BibitemOpen [0]{}%
\providecommand \bibitemStop [0]{}%
\providecommand \bibitemNoStop [0]{.\EOS\space}%
\providecommand \EOS [0]{\spacefactor3000\relax}%
\providecommand \BibitemShut  [1]{\csname bibitem#1\endcsname}%
\let\auto@bib@innerbib\@empty
\bibitem [{\citenamefont {{Barthel}}\ \emph {et~al.}(2009)\citenamefont
  {{Barthel}}, \citenamefont {{Kasztelan}}, \citenamefont {{McCulloch}},\ and\
  \citenamefont {{Schollw{\"o}ck}}}]{barthel2009magnetism}%
  \BibitemOpen
  \bibfield  {author} {\bibinfo {author} {\bibfnamefont {T.}~\bibnamefont
  {{Barthel}}}, \bibinfo {author} {\bibfnamefont {C.}~\bibnamefont
  {{Kasztelan}}}, \bibinfo {author} {\bibfnamefont {I.~P.}\ \bibnamefont
  {{McCulloch}}}, \ and\ \bibinfo {author} {\bibfnamefont {U.}~\bibnamefont
  {{Schollw{\"o}ck}}},\ }\href {\doibase 10.1103/PhysRevA.79.053627} {\bibfield
   {journal} {\bibinfo  {journal} {\pra}\ }\textbf {\bibinfo {volume} {79}},\
  \bibinfo {eid} {053627} (\bibinfo {year} {2009})},\ \Eprint
  {http://arxiv.org/abs/0809.5141} {arXiv:0809.5141 [cond-mat.stat-mech]}
  \BibitemShut {NoStop}%
\bibitem [{\citenamefont {{Schrieffer}}\ and\ \citenamefont
  {{Wolff}}(1966)}]{schrieffer1966relation}%
  \BibitemOpen
  \bibfield  {author} {\bibinfo {author} {\bibfnamefont {J.~R.}\ \bibnamefont
  {{Schrieffer}}}\ and\ \bibinfo {author} {\bibfnamefont {P.~A.}\ \bibnamefont
  {{Wolff}}},\ }\href {\doibase 10.1103/PhysRev.149.491} {\bibfield  {journal}
  {\bibinfo  {journal} {Physical Review}\ }\textbf {\bibinfo {volume} {149}},\
  \bibinfo {pages} {491} (\bibinfo {year} {1966})}\BibitemShut {NoStop}%
\bibitem [{\citenamefont {{Zhu}}\ \emph {et~al.}(2014)\citenamefont {{Zhu}},
  \citenamefont {{Li}},\ and\ \citenamefont {{Batista}}}]{zhu2014spin}%
  \BibitemOpen
  \bibfield  {author} {\bibinfo {author} {\bibfnamefont {S.}~\bibnamefont
  {{Zhu}}}, \bibinfo {author} {\bibfnamefont {Y.-Q.}\ \bibnamefont {{Li}}}, \
  and\ \bibinfo {author} {\bibfnamefont {C.~D.}\ \bibnamefont {{Batista}}},\
  }\href {\doibase 10.1103/PhysRevB.90.195107} {\bibfield  {journal} {\bibinfo
  {journal} {\prb}\ }\textbf {\bibinfo {volume} {90}},\ \bibinfo {eid} {195107}
  (\bibinfo {year} {2014})},\ \Eprint {http://arxiv.org/abs/1411.2511}
  {arXiv:1411.2511 [cond-mat.str-el]} \BibitemShut {NoStop}%
\bibitem [{\citenamefont {{Pixley}}\ \emph {et~al.}(2017)\citenamefont
  {{Pixley}}, \citenamefont {{Cole}}, \citenamefont {{Spielman}}, \citenamefont
  {{Rizzi}},\ and\ \citenamefont {{Das Sarma}}}]{pixley2017strong}%
  \BibitemOpen
  \bibfield  {author} {\bibinfo {author} {\bibfnamefont {J.~H.}\ \bibnamefont
  {{Pixley}}}, \bibinfo {author} {\bibfnamefont {W.~S.}\ \bibnamefont
  {{Cole}}}, \bibinfo {author} {\bibfnamefont {I.~B.}\ \bibnamefont
  {{Spielman}}}, \bibinfo {author} {\bibfnamefont {M.}~\bibnamefont {{Rizzi}}},
  \ and\ \bibinfo {author} {\bibfnamefont {S.}~\bibnamefont {{Das Sarma}}},\
  }\href {\doibase 10.1103/PhysRevA.96.043622} {\bibfield  {journal} {\bibinfo
  {journal} {\pra}\ }\textbf {\bibinfo {volume} {96}},\ \bibinfo {eid} {043622}
  (\bibinfo {year} {2017})},\ \Eprint {http://arxiv.org/abs/1705.06742}
  {arXiv:1705.06742 [cond-mat.quant-gas]} \BibitemShut {NoStop}%
\end{thebibliography}%


\begin{thebibliography}{81}%
\makeatletter
\providecommand \@ifxundefined [1]{%
 \@ifx{#1\undefined}
}%
\providecommand \@ifnum [1]{%
 \ifnum #1\expandafter \@firstoftwo
 \else \expandafter \@secondoftwo
 \fi
}%
\providecommand \@ifx [1]{%
 \ifx #1\expandafter \@firstoftwo
 \else \expandafter \@secondoftwo
 \fi
}%
\providecommand \natexlab [1]{#1}%
\providecommand \enquote  [1]{``#1''}%
\providecommand \bibnamefont  [1]{#1}%
\providecommand \bibfnamefont [1]{#1}%
\providecommand \citenamefont [1]{#1}%
\providecommand \href@noop [0]{\@secondoftwo}%
\providecommand \href [0]{\begingroup \@sanitize@url \@href}%
\providecommand \@href[1]{\@@startlink{#1}\@@href}%
\providecommand \@@href[1]{\endgroup#1\@@endlink}%
\providecommand \@sanitize@url [0]{\catcode `\\12\catcode `\$12\catcode
  `\&12\catcode `\#12\catcode `\^12\catcode `\_12\catcode `\%12\relax}%
\providecommand \@@startlink[1]{}%
\providecommand \@@endlink[0]{}%
\providecommand \url  [0]{\begingroup\@sanitize@url \@url }%
\providecommand \@url [1]{\endgroup\@href {#1}{\urlprefix }}%
\providecommand \urlprefix  [0]{URL }%
\providecommand \Eprint [0]{\href }%
\providecommand \doibase [0]{http://dx.doi.org/}%
\providecommand \selectlanguage [0]{\@gobble}%
\providecommand \bibinfo  [0]{\@secondoftwo}%
\providecommand \bibfield  [0]{\@secondoftwo}%
\providecommand \translation [1]{[#1]}%
\providecommand \BibitemOpen [0]{}%
\providecommand \bibitemStop [0]{}%
\providecommand \bibitemNoStop [0]{.\EOS\space}%
\providecommand \EOS [0]{\spacefactor3000\relax}%
\providecommand \BibitemShut  [1]{\csname bibitem#1\endcsname}%
\let\auto@bib@innerbib\@empty
\bibitem [{\citenamefont {Skyrme}(1962)}]{Skyrme:1962vh}%
  \BibitemOpen
  \bibfield  {author} {\bibinfo {author} {\bibfnamefont {T.~H.~R.}\
  \bibnamefont {Skyrme}},\ }\href {\doibase 10.1016/0029-5582(62)90775-7}
  {\bibfield  {journal} {\bibinfo  {journal} {Nucl. Phys.}\ }\textbf {\bibinfo
  {volume} {31}},\ \bibinfo {pages} {556} (\bibinfo {year} {1962})}\BibitemShut
  {NoStop}%
\bibitem [{\citenamefont {Witten}(1983)}]{Witten:1983tx}%
  \BibitemOpen
  \bibfield  {author} {\bibinfo {author} {\bibfnamefont {E.}~\bibnamefont
  {Witten}},\ }\href {\doibase 10.1016/0550-3213(83)90064-0} {\bibfield
  {journal} {\bibinfo  {journal} {Nucl. Phys. B}\ }\textbf {\bibinfo {volume}
  {223}},\ \bibinfo {pages} {433} (\bibinfo {year} {1983})}\BibitemShut
  {NoStop}%
\bibitem [{\citenamefont {Manton}(2022)}]{MantonsBook}%
  \BibitemOpen
  \bibfield  {author} {\bibinfo {author} {\bibfnamefont {N.}~\bibnamefont
  {Manton}},\ }\href {\doibase 10.1142/q0368} {\emph {\bibinfo {title}
  {{Skyrmions - A Theory of Nuclei}}}}\ (\bibinfo  {publisher} {World
  Scientific},\ \bibinfo {address} {Singapore},\ \bibinfo {year}
  {2022})\BibitemShut {NoStop}%
\bibitem [{\citenamefont {Rho}\ and\ \citenamefont
  {Zahed}(2016{\natexlab{a}})}]{multifaceted}%
  \BibitemOpen
  \bibinfo {editor} {\bibfnamefont {M.}~\bibnamefont {Rho}}\ and\ \bibinfo
  {editor} {\bibfnamefont {I.}~\bibnamefont {Zahed}},\ eds.,\ \href@noop {}
  {\emph {\bibinfo {title} {{The Multifaceted Skyrmions}}}},\ \bibinfo
  {edition} {2nd}\ ed.\ (\bibinfo  {publisher} {World Scientific},\ \bibinfo
  {address} {Singapore},\ \bibinfo {year} {2016})\BibitemShut {NoStop}%
\bibitem [{\citenamefont {Bogolubskaya}\ and\ \citenamefont
  {Bogolubsky}(1989)}]{Bogolubskaya:1989ha}%
  \BibitemOpen
  \bibfield  {author} {\bibinfo {author} {\bibfnamefont {A.~A.}\ \bibnamefont
  {Bogolubskaya}}\ and\ \bibinfo {author} {\bibfnamefont {I.~L.}\ \bibnamefont
  {Bogolubsky}},\ }\href {\doibase 10.1016/0375-9601(89)90301-0} {\bibfield
  {journal} {\bibinfo  {journal} {Phys. Lett. A}\ }\textbf {\bibinfo {volume}
  {136}},\ \bibinfo {pages} {485} (\bibinfo {year} {1989})}\BibitemShut
  {NoStop}%
\bibitem [{\citenamefont {Bogolubskaya}\ and\ \citenamefont
  {Bogolubsky}(1990)}]{doi:10.1007/BF01045888}%
  \BibitemOpen
  \bibfield  {author} {\bibinfo {author} {\bibfnamefont {A.~A.}\ \bibnamefont
  {Bogolubskaya}}\ and\ \bibinfo {author} {\bibfnamefont {I.~L.}\ \bibnamefont
  {Bogolubsky}},\ }\href {\doibase 10.1007/BF01045888} {\bibfield  {journal}
  {\bibinfo  {journal} {Letters in Mathematical Physics}\ }\textbf {\bibinfo
  {volume} {19}},\ \bibinfo {pages} {171} (\bibinfo {year} {1990})}\BibitemShut
  {NoStop}%
\bibitem [{\citenamefont {Piette}\ \emph {et~al.}(1995)\citenamefont {Piette},
  \citenamefont {Schroers},\ and\ \citenamefont {Zakrzewski}}]{Piette:1994ug}%
  \BibitemOpen
  \bibfield  {author} {\bibinfo {author} {\bibfnamefont {B.~M. A.~G.}\
  \bibnamefont {Piette}}, \bibinfo {author} {\bibfnamefont {B.~J.}\
  \bibnamefont {Schroers}}, \ and\ \bibinfo {author} {\bibfnamefont {W.~J.}\
  \bibnamefont {Zakrzewski}},\ }\href {\doibase 10.1007/BF01571317} {\bibfield
  {journal} {\bibinfo  {journal} {Z. Phys. C}\ }\textbf {\bibinfo {volume}
  {65}},\ \bibinfo {pages} {165} (\bibinfo {year} {1995})},\ \Eprint
  {http://arxiv.org/abs/hep-th/9406160} {arXiv:hep-th/9406160} \BibitemShut
  {NoStop}%
\bibitem [{\citenamefont {Bogdanov}\ and\ \citenamefont
  {Yablonskii}(1989)}]{Bogdanov:1989}%
  \BibitemOpen
  \bibfield  {author} {\bibinfo {author} {\bibfnamefont {A.}~\bibnamefont
  {Bogdanov}}\ and\ \bibinfo {author} {\bibfnamefont {D.}~\bibnamefont
  {Yablonskii}},\ }\href@noop {} {\bibfield  {journal} {\bibinfo  {journal}
  {Sov. Phys. JETP}\ }\textbf {\bibinfo {volume} {68}},\ \bibinfo {pages} {101}
  (\bibinfo {year} {1989})}\BibitemShut {NoStop}%
\bibitem [{\citenamefont {Bogdanov}(1995)}]{Bogdanov:1995}%
  \BibitemOpen
  \bibfield  {author} {\bibinfo {author} {\bibfnamefont {A.}~\bibnamefont
  {Bogdanov}},\ }\href@noop {} {\bibfield  {journal} {\bibinfo  {journal} {JETP
  Lett.}\ }\textbf {\bibinfo {volume} {62}},\ \bibinfo {pages} {247} (\bibinfo
  {year} {1995})}\BibitemShut {NoStop}%
\bibitem [{\citenamefont {Mühlbauer}\ \emph {et~al.}(2009)\citenamefont
  {Mühlbauer}, \citenamefont {Binz}, \citenamefont {Jonietz}, \citenamefont
  {Pfleiderer}, \citenamefont {Rosch}, \citenamefont {Neubauer}, \citenamefont
  {Georgii},\ and\ \citenamefont {Böni}}]{doi:10.1126/science.1166767}%
  \BibitemOpen
  \bibfield  {author} {\bibinfo {author} {\bibfnamefont {S.}~\bibnamefont
  {Mühlbauer}}, \bibinfo {author} {\bibfnamefont {B.}~\bibnamefont {Binz}},
  \bibinfo {author} {\bibfnamefont {F.}~\bibnamefont {Jonietz}}, \bibinfo
  {author} {\bibfnamefont {C.}~\bibnamefont {Pfleiderer}}, \bibinfo {author}
  {\bibfnamefont {A.}~\bibnamefont {Rosch}}, \bibinfo {author} {\bibfnamefont
  {A.}~\bibnamefont {Neubauer}}, \bibinfo {author} {\bibfnamefont
  {R.}~\bibnamefont {Georgii}}, \ and\ \bibinfo {author} {\bibfnamefont
  {P.}~\bibnamefont {Böni}},\ }\href {\doibase 10.1126/science.1166767}
  {\bibfield  {journal} {\bibinfo  {journal} {Science}\ }\textbf {\bibinfo
  {volume} {323}},\ \bibinfo {pages} {915} (\bibinfo {year}
  {2009})}\BibitemShut {NoStop}%
\bibitem [{\citenamefont {Yu}\ \emph {et~al.}(2010)\citenamefont {Yu},
  \citenamefont {Onose}, \citenamefont {Kanazawa}, \citenamefont {Park},
  \citenamefont {Han}, \citenamefont {Matsui}, \citenamefont {Nagaosa},\ and\
  \citenamefont {Tokura}}]{doi:10.1038/nature09124}%
  \BibitemOpen
  \bibfield  {author} {\bibinfo {author} {\bibfnamefont {X.~Z.}\ \bibnamefont
  {Yu}}, \bibinfo {author} {\bibfnamefont {Y.}~\bibnamefont {Onose}}, \bibinfo
  {author} {\bibfnamefont {N.}~\bibnamefont {Kanazawa}}, \bibinfo {author}
  {\bibfnamefont {J.~H.}\ \bibnamefont {Park}}, \bibinfo {author}
  {\bibfnamefont {J.~H.}\ \bibnamefont {Han}}, \bibinfo {author} {\bibfnamefont
  {Y.}~\bibnamefont {Matsui}}, \bibinfo {author} {\bibfnamefont
  {N.}~\bibnamefont {Nagaosa}}, \ and\ \bibinfo {author} {\bibfnamefont
  {Y.}~\bibnamefont {Tokura}},\ }\href {\doibase 10.1038/nature09124}
  {\bibfield  {journal} {\bibinfo  {journal} {Nature}\ }\textbf {\bibinfo
  {volume} {465}},\ \bibinfo {pages} {901} (\bibinfo {year}
  {2010})}\BibitemShut {NoStop}%
\bibitem [{\citenamefont {{Heinze}}\ \emph {et~al.}(2011)\citenamefont
  {{Heinze}}, \citenamefont {{von Bergmann}}, \citenamefont {{Menzel}},
  \citenamefont {{Brede}}, \citenamefont {{Kubetzka}}, \citenamefont
  {{Wiesendanger}}, \citenamefont {{Bihlmayer}},\ and\ \citenamefont
  {{Bl{\"u}gel}}}]{doi:10.1038/nphys2045}%
  \BibitemOpen
  \bibfield  {author} {\bibinfo {author} {\bibfnamefont {S.}~\bibnamefont
  {{Heinze}}}, \bibinfo {author} {\bibfnamefont {K.}~\bibnamefont {{von
  Bergmann}}}, \bibinfo {author} {\bibfnamefont {M.}~\bibnamefont {{Menzel}}},
  \bibinfo {author} {\bibfnamefont {J.}~\bibnamefont {{Brede}}}, \bibinfo
  {author} {\bibfnamefont {A.}~\bibnamefont {{Kubetzka}}}, \bibinfo {author}
  {\bibfnamefont {R.}~\bibnamefont {{Wiesendanger}}}, \bibinfo {author}
  {\bibfnamefont {G.}~\bibnamefont {{Bihlmayer}}}, \ and\ \bibinfo {author}
  {\bibfnamefont {S.}~\bibnamefont {{Bl{\"u}gel}}},\ }\href {\doibase
  https://doi.org/10.1038/nphys2045} {\bibfield  {journal} {\bibinfo  {journal}
  {Nature Physics}\ }\textbf {\bibinfo {volume} {7}},\ \bibinfo {pages} {713}
  (\bibinfo {year} {2011})}\BibitemShut {NoStop}%
\bibitem [{\citenamefont {Kurumaji}\ \emph {et~al.}(2019)\citenamefont
  {Kurumaji}, \citenamefont {Nakajima}, \citenamefont {Hirschberger},
  \citenamefont {Kikkawa}, \citenamefont {Yamasaki}, \citenamefont {Sagayama},
  \citenamefont {Nakao}, \citenamefont {Taguchi}, \citenamefont {Arima},\ and\
  \citenamefont {Tokura}}]{Kurumaji2019}%
  \BibitemOpen
  \bibfield  {author} {\bibinfo {author} {\bibfnamefont {T.}~\bibnamefont
  {Kurumaji}}, \bibinfo {author} {\bibfnamefont {T.}~\bibnamefont {Nakajima}},
  \bibinfo {author} {\bibfnamefont {M.}~\bibnamefont {Hirschberger}}, \bibinfo
  {author} {\bibfnamefont {A.}~\bibnamefont {Kikkawa}}, \bibinfo {author}
  {\bibfnamefont {Y.}~\bibnamefont {Yamasaki}}, \bibinfo {author}
  {\bibfnamefont {H.}~\bibnamefont {Sagayama}}, \bibinfo {author}
  {\bibfnamefont {H.}~\bibnamefont {Nakao}}, \bibinfo {author} {\bibfnamefont
  {Y.}~\bibnamefont {Taguchi}}, \bibinfo {author} {\bibfnamefont {T.-h.}\
  \bibnamefont {Arima}}, \ and\ \bibinfo {author} {\bibfnamefont
  {Y.}~\bibnamefont {Tokura}},\ }\href {\doibase 10.1126/science.aau0968}
  {\bibfield  {journal} {\bibinfo  {journal} {Science}\ }\textbf {\bibinfo
  {volume} {365}},\ \bibinfo {pages} {914} (\bibinfo {year}
  {2019})}\BibitemShut {NoStop}%
\bibitem [{\citenamefont {Hirschberger}\ \emph {et~al.}(2019)\citenamefont
  {Hirschberger}, \citenamefont {Nakajima}, \citenamefont {Gao}, \citenamefont
  {Peng}, \citenamefont {Kikkawa}, \citenamefont {Kurumaji}, \citenamefont
  {Kriener}, \citenamefont {Yamasaki}, \citenamefont {Sagayama}, \citenamefont
  {Nakao}, \citenamefont {Ohishi}, \citenamefont {Kakurai}, \citenamefont
  {Taguchi}, \citenamefont {Yu}, \citenamefont {Arima},\ and\ \citenamefont
  {Tokura}}]{Hirschberger2019}%
  \BibitemOpen
  \bibfield  {author} {\bibinfo {author} {\bibfnamefont {M.}~\bibnamefont
  {Hirschberger}}, \bibinfo {author} {\bibfnamefont {T.}~\bibnamefont
  {Nakajima}}, \bibinfo {author} {\bibfnamefont {S.}~\bibnamefont {Gao}},
  \bibinfo {author} {\bibfnamefont {L.}~\bibnamefont {Peng}}, \bibinfo {author}
  {\bibfnamefont {A.}~\bibnamefont {Kikkawa}}, \bibinfo {author} {\bibfnamefont
  {T.}~\bibnamefont {Kurumaji}}, \bibinfo {author} {\bibfnamefont
  {M.}~\bibnamefont {Kriener}}, \bibinfo {author} {\bibfnamefont
  {Y.}~\bibnamefont {Yamasaki}}, \bibinfo {author} {\bibfnamefont
  {H.}~\bibnamefont {Sagayama}}, \bibinfo {author} {\bibfnamefont
  {H.}~\bibnamefont {Nakao}}, \bibinfo {author} {\bibfnamefont
  {K.}~\bibnamefont {Ohishi}}, \bibinfo {author} {\bibfnamefont
  {K.}~\bibnamefont {Kakurai}}, \bibinfo {author} {\bibfnamefont
  {Y.}~\bibnamefont {Taguchi}}, \bibinfo {author} {\bibfnamefont
  {X.}~\bibnamefont {Yu}}, \bibinfo {author} {\bibfnamefont {T.-h.}\
  \bibnamefont {Arima}}, \ and\ \bibinfo {author} {\bibfnamefont
  {Y.}~\bibnamefont {Tokura}},\ }\href {\doibase 10.1038/s41467-019-13675-4}
  {\bibfield  {journal} {\bibinfo  {journal} {Nat. Commun.}\ }\textbf {\bibinfo
  {volume} {10}},\ \bibinfo {pages} {5831} (\bibinfo {year}
  {2019})}\BibitemShut {NoStop}%
\bibitem [{\citenamefont {Khanh}\ \emph {et~al.}(2020)\citenamefont {Khanh},
  \citenamefont {Nakajima}, \citenamefont {Yu}, \citenamefont {Gao},
  \citenamefont {Shibata}, \citenamefont {Hirschberger}, \citenamefont
  {Yamasaki}, \citenamefont {Sagayama}, \citenamefont {Nakao}, \citenamefont
  {Peng}, \citenamefont {Nakajima}, \citenamefont {Takagi}, \citenamefont
  {Arima}, \citenamefont {Tokura},\ and\ \citenamefont {Seki}}]{Khanh2020}%
  \BibitemOpen
  \bibfield  {author} {\bibinfo {author} {\bibfnamefont {N.~D.}\ \bibnamefont
  {Khanh}}, \bibinfo {author} {\bibfnamefont {T.}~\bibnamefont {Nakajima}},
  \bibinfo {author} {\bibfnamefont {X.}~\bibnamefont {Yu}}, \bibinfo {author}
  {\bibfnamefont {S.}~\bibnamefont {Gao}}, \bibinfo {author} {\bibfnamefont
  {K.}~\bibnamefont {Shibata}}, \bibinfo {author} {\bibfnamefont
  {M.}~\bibnamefont {Hirschberger}}, \bibinfo {author} {\bibfnamefont
  {Y.}~\bibnamefont {Yamasaki}}, \bibinfo {author} {\bibfnamefont
  {H.}~\bibnamefont {Sagayama}}, \bibinfo {author} {\bibfnamefont
  {H.}~\bibnamefont {Nakao}}, \bibinfo {author} {\bibfnamefont
  {L.}~\bibnamefont {Peng}}, \bibinfo {author} {\bibfnamefont {K.}~\bibnamefont
  {Nakajima}}, \bibinfo {author} {\bibfnamefont {R.}~\bibnamefont {Takagi}},
  \bibinfo {author} {\bibfnamefont {T.-h.}\ \bibnamefont {Arima}}, \bibinfo
  {author} {\bibfnamefont {Y.}~\bibnamefont {Tokura}}, \ and\ \bibinfo {author}
  {\bibfnamefont {S.}~\bibnamefont {Seki}},\ }\href {\doibase
  https://doi.org/10.1038/s41565-020-0684-7} {\bibfield  {journal} {\bibinfo
  {journal} {Nat. Nanotechnol.}\ }\textbf {\bibinfo {volume} {15}},\ \bibinfo
  {pages} {444} (\bibinfo {year} {2020})}\BibitemShut {NoStop}%
\bibitem [{\citenamefont {Yasui}\ \emph {et~al.}(2020)\citenamefont {Yasui},
  \citenamefont {Butler}, \citenamefont {Khanh}, \citenamefont {Hayami},
  \citenamefont {Nomoto}, \citenamefont {Hanaguri}, \citenamefont {Motome},
  \citenamefont {Arita}, \citenamefont {Arima}, \citenamefont {Tokura},\ and\
  \citenamefont {Seki}}]{Yasui2020}%
  \BibitemOpen
  \bibfield  {author} {\bibinfo {author} {\bibfnamefont {Y.}~\bibnamefont
  {Yasui}}, \bibinfo {author} {\bibfnamefont {C.~J.}\ \bibnamefont {Butler}},
  \bibinfo {author} {\bibfnamefont {N.~D.}\ \bibnamefont {Khanh}}, \bibinfo
  {author} {\bibfnamefont {S.}~\bibnamefont {Hayami}}, \bibinfo {author}
  {\bibfnamefont {T.}~\bibnamefont {Nomoto}}, \bibinfo {author} {\bibfnamefont
  {T.}~\bibnamefont {Hanaguri}}, \bibinfo {author} {\bibfnamefont
  {Y.}~\bibnamefont {Motome}}, \bibinfo {author} {\bibfnamefont
  {R.}~\bibnamefont {Arita}}, \bibinfo {author} {\bibfnamefont {T.-h.}\
  \bibnamefont {Arima}}, \bibinfo {author} {\bibfnamefont {Y.}~\bibnamefont
  {Tokura}}, \ and\ \bibinfo {author} {\bibfnamefont {S.}~\bibnamefont
  {Seki}},\ }\href {\doibase 10.1038/s41467-020-19751-4} {\bibfield  {journal}
  {\bibinfo  {journal} {Nat. Commun.}\ }\textbf {\bibinfo {volume} {11}},\
  \bibinfo {pages} {5925} (\bibinfo {year} {2020})}\BibitemShut {NoStop}%
\bibitem [{\citenamefont {Fert}\ \emph {et~al.}(2013)\citenamefont {Fert},
  \citenamefont {Cros},\ and\ \citenamefont
  {Sampaio}}]{doi:10.1038/nnano.2013.29}%
  \BibitemOpen
  \bibfield  {author} {\bibinfo {author} {\bibfnamefont {A.}~\bibnamefont
  {Fert}}, \bibinfo {author} {\bibfnamefont {V.}~\bibnamefont {Cros}}, \ and\
  \bibinfo {author} {\bibfnamefont {J.}~\bibnamefont {Sampaio}},\ }\href
  {\doibase 10.1038/nnano.2013.29} {\bibfield  {journal} {\bibinfo  {journal}
  {Nature Nanotechnology}\ }\textbf {\bibinfo {volume} {8}},\ \bibinfo {pages}
  {152} (\bibinfo {year} {2013})}\BibitemShut {NoStop}%
\bibitem [{\citenamefont {Nagaosa}\ and\ \citenamefont
  {Tokura}(2013)}]{Nagaosa2013}%
  \BibitemOpen
  \bibfield  {author} {\bibinfo {author} {\bibfnamefont {N.}~\bibnamefont
  {Nagaosa}}\ and\ \bibinfo {author} {\bibfnamefont {Y.}~\bibnamefont
  {Tokura}},\ }\href {\doibase 10.1038/nnano.2013.243} {\bibfield  {journal}
  {\bibinfo  {journal} {Nature Nanotech}\ }\textbf {\bibinfo {volume} {8}},\
  \bibinfo {pages} {899–911} (\bibinfo {year} {2013})}\BibitemShut {NoStop}%
\bibitem [{\citenamefont {Okubo}\ \emph {et~al.}(2012)\citenamefont {Okubo},
  \citenamefont {Chung},\ and\ \citenamefont
  {Kawamura}}]{PhysRevLett.108.017206}%
  \BibitemOpen
  \bibfield  {author} {\bibinfo {author} {\bibfnamefont {T.}~\bibnamefont
  {Okubo}}, \bibinfo {author} {\bibfnamefont {S.}~\bibnamefont {Chung}}, \ and\
  \bibinfo {author} {\bibfnamefont {H.}~\bibnamefont {Kawamura}},\ }\href
  {\doibase 10.1103/PhysRevLett.108.017206} {\bibfield  {journal} {\bibinfo
  {journal} {Phys. Rev. Lett.}\ }\textbf {\bibinfo {volume} {108}},\ \bibinfo
  {pages} {017206} (\bibinfo {year} {2012})}\BibitemShut {NoStop}%
\bibitem [{\citenamefont {Leonov}\ and\ \citenamefont
  {Mostovoy}(2015)}]{Leonov2015}%
  \BibitemOpen
  \bibfield  {author} {\bibinfo {author} {\bibfnamefont {A.~O.}\ \bibnamefont
  {Leonov}}\ and\ \bibinfo {author} {\bibfnamefont {M.}~\bibnamefont
  {Mostovoy}},\ }\href {\doibase 10.1038/ncomms9275} {\bibfield  {journal}
  {\bibinfo  {journal} {Nat. Commun.}\ }\textbf {\bibinfo {volume} {6}},\
  \bibinfo {pages} {8275} (\bibinfo {year} {2015})}\BibitemShut {NoStop}%
\bibitem [{\citenamefont {Amoroso}\ \emph {et~al.}(2020)\citenamefont
  {Amoroso}, \citenamefont {Barone},\ and\ \citenamefont
  {Picozzi}}]{Amoroso2020}%
  \BibitemOpen
  \bibfield  {author} {\bibinfo {author} {\bibfnamefont {D.}~\bibnamefont
  {Amoroso}}, \bibinfo {author} {\bibfnamefont {P.}~\bibnamefont {Barone}}, \
  and\ \bibinfo {author} {\bibfnamefont {S.}~\bibnamefont {Picozzi}},\ }\href
  {\doibase 10.1038/s41467-020-19535-w} {\bibfield  {journal} {\bibinfo
  {journal} {Nat. Commun.}\ }\textbf {\bibinfo {volume} {11}},\ \bibinfo
  {pages} {5784} (\bibinfo {year} {2020})}\BibitemShut {NoStop}%
\bibitem [{\citenamefont {Akagi}\ \emph {et~al.}(2012)\citenamefont {Akagi},
  \citenamefont {Udagawa},\ and\ \citenamefont {Motome}}]{Akagi2012}%
  \BibitemOpen
  \bibfield  {author} {\bibinfo {author} {\bibfnamefont {Y.}~\bibnamefont
  {Akagi}}, \bibinfo {author} {\bibfnamefont {M.}~\bibnamefont {Udagawa}}, \
  and\ \bibinfo {author} {\bibfnamefont {Y.}~\bibnamefont {Motome}},\ }\href
  {\doibase 10.1103/PhysRevLett.108.096401} {\bibfield  {journal} {\bibinfo
  {journal} {Phys. Rev. Lett.}\ }\textbf {\bibinfo {volume} {108}},\ \bibinfo
  {pages} {096401} (\bibinfo {year} {2012})}\BibitemShut {NoStop}%
\bibitem [{\citenamefont {Ozawa}\ \emph {et~al.}(2017)\citenamefont {Ozawa},
  \citenamefont {Hayami},\ and\ \citenamefont {Motome}}]{Ozawa2017}%
  \BibitemOpen
  \bibfield  {author} {\bibinfo {author} {\bibfnamefont {R.}~\bibnamefont
  {Ozawa}}, \bibinfo {author} {\bibfnamefont {S.}~\bibnamefont {Hayami}}, \
  and\ \bibinfo {author} {\bibfnamefont {Y.}~\bibnamefont {Motome}},\ }\href
  {\doibase 10.1103/PhysRevLett.118.147205} {\bibfield  {journal} {\bibinfo
  {journal} {Phys. Rev. Lett.}\ }\textbf {\bibinfo {volume} {118}},\ \bibinfo
  {pages} {147205} (\bibinfo {year} {2017})}\BibitemShut {NoStop}%
\bibitem [{\citenamefont {Hayami}\ \emph {et~al.}(2017)\citenamefont {Hayami},
  \citenamefont {Ozawa},\ and\ \citenamefont {Motome}}]{Hayami2017a}%
  \BibitemOpen
  \bibfield  {author} {\bibinfo {author} {\bibfnamefont {S.}~\bibnamefont
  {Hayami}}, \bibinfo {author} {\bibfnamefont {R.}~\bibnamefont {Ozawa}}, \
  and\ \bibinfo {author} {\bibfnamefont {Y.}~\bibnamefont {Motome}},\ }\href
  {\doibase 10.1103/PhysRevB.95.224424} {\bibfield  {journal} {\bibinfo
  {journal} {Phys. Rev. B}\ }\textbf {\bibinfo {volume} {95}},\ \bibinfo
  {pages} {224424} (\bibinfo {year} {2017})}\BibitemShut {NoStop}%
\bibitem [{\citenamefont {Eichenherr}(1978)}]{Eichenherr:1978qa}%
  \BibitemOpen
  \bibfield  {author} {\bibinfo {author} {\bibfnamefont {H.}~\bibnamefont
  {Eichenherr}},\ }\href {\doibase 10.1016/0550-3213(78)90439-X} {\bibfield
  {journal} {\bibinfo  {journal} {Nucl. Phys. B}\ }\textbf {\bibinfo {volume}
  {146}},\ \bibinfo {pages} {215} (\bibinfo {year} {1978})},\ \bibinfo {note}
  {[Erratum: Nucl.Phys.B 155, 544 (1979)]}\BibitemShut {NoStop}%
\bibitem [{\citenamefont {Golo}\ and\ \citenamefont
  {Perelomov}(1978)}]{Golo:1978de}%
  \BibitemOpen
  \bibfield  {author} {\bibinfo {author} {\bibfnamefont {V.~L.}\ \bibnamefont
  {Golo}}\ and\ \bibinfo {author} {\bibfnamefont {A.~M.}\ \bibnamefont
  {Perelomov}},\ }\href {\doibase 10.1016/0370-2693(78)90447-1} {\bibfield
  {journal} {\bibinfo  {journal} {Phys. Lett. B}\ }\textbf {\bibinfo {volume}
  {79}},\ \bibinfo {pages} {112} (\bibinfo {year} {1978})}\BibitemShut
  {NoStop}%
\bibitem [{\citenamefont {Cremmer}\ and\ \citenamefont
  {Scherk}(1978)}]{Cremmer:1978bh}%
  \BibitemOpen
  \bibfield  {author} {\bibinfo {author} {\bibfnamefont {E.}~\bibnamefont
  {Cremmer}}\ and\ \bibinfo {author} {\bibfnamefont {J.}~\bibnamefont
  {Scherk}},\ }\href {\doibase 10.1016/0370-2693(78)90672-X} {\bibfield
  {journal} {\bibinfo  {journal} {Phys. Lett. B}\ }\textbf {\bibinfo {volume}
  {74}},\ \bibinfo {pages} {341} (\bibinfo {year} {1978})}\BibitemShut
  {NoStop}%
\bibitem [{\citenamefont {D'Adda}\ \emph {et~al.}(1978)\citenamefont {D'Adda},
  \citenamefont {Luscher},\ and\ \citenamefont {Di~Vecchia}}]{DAdda:1978vbw}%
  \BibitemOpen
  \bibfield  {author} {\bibinfo {author} {\bibfnamefont {A.}~\bibnamefont
  {D'Adda}}, \bibinfo {author} {\bibfnamefont {M.}~\bibnamefont {Luscher}}, \
  and\ \bibinfo {author} {\bibfnamefont {P.}~\bibnamefont {Di~Vecchia}},\
  }\href {\doibase 10.1016/0550-3213(78)90432-7} {\bibfield  {journal}
  {\bibinfo  {journal} {Nucl. Phys. B}\ }\textbf {\bibinfo {volume} {146}},\
  \bibinfo {pages} {63} (\bibinfo {year} {1978})}\BibitemShut {NoStop}%
\bibitem [{\citenamefont {Witten}(1979)}]{Witten:1978bc}%
  \BibitemOpen
  \bibfield  {author} {\bibinfo {author} {\bibfnamefont {E.}~\bibnamefont
  {Witten}},\ }\href {\doibase 10.1016/0550-3213(79)90243-8} {\bibfield
  {journal} {\bibinfo  {journal} {Nucl. Phys. B}\ }\textbf {\bibinfo {volume}
  {149}},\ \bibinfo {pages} {285} (\bibinfo {year} {1979})}\BibitemShut
  {NoStop}%
\bibitem [{\citenamefont {Senthil}\ \emph {et~al.}(2004)\citenamefont
  {Senthil}, \citenamefont {Vishwanath}, \citenamefont {Balents}, \citenamefont
  {Sachdev},\ and\ \citenamefont {Fisher}}]{Senthil:2003eed}%
  \BibitemOpen
  \bibfield  {author} {\bibinfo {author} {\bibfnamefont {T.}~\bibnamefont
  {Senthil}}, \bibinfo {author} {\bibfnamefont {A.}~\bibnamefont {Vishwanath}},
  \bibinfo {author} {\bibfnamefont {L.}~\bibnamefont {Balents}}, \bibinfo
  {author} {\bibfnamefont {S.}~\bibnamefont {Sachdev}}, \ and\ \bibinfo
  {author} {\bibfnamefont {M.~P.~A.}\ \bibnamefont {Fisher}},\ }\href {\doibase
  https://doi.org/10.1126/science.1091806} {\bibfield  {journal} {\bibinfo
  {journal} {Science}\ }\textbf {\bibinfo {volume} {303}},\ \bibinfo {pages}
  {1490} (\bibinfo {year} {2004})},\ \Eprint
  {http://arxiv.org/abs/cond-mat/0311326} {arXiv:cond-mat/0311326} \BibitemShut
  {NoStop}%
\bibitem [{\citenamefont {Nogueira}\ and\ \citenamefont
  {Sudb\o{}}(2013)}]{Nogueira:2013oza}%
  \BibitemOpen
  \bibfield  {author} {\bibinfo {author} {\bibfnamefont {F.~S.}\ \bibnamefont
  {Nogueira}}\ and\ \bibinfo {author} {\bibfnamefont {A.}~\bibnamefont
  {Sudb\o{}}},\ }\href {\doibase 10.1209/0295-5075/104/56004} {\bibfield
  {journal} {\bibinfo  {journal} {EPL}\ }\textbf {\bibinfo {volume} {104}},\
  \bibinfo {pages} {56004} (\bibinfo {year} {2013})},\ \Eprint
  {http://arxiv.org/abs/1304.4938} {arXiv:1304.4938 [cond-mat.str-el]}
  \BibitemShut {NoStop}%
\bibitem [{\citenamefont {Laflamme}\ \emph {et~al.}(2016)\citenamefont
  {Laflamme}, \citenamefont {Evans}, \citenamefont {Dalmonte}, \citenamefont
  {Gerber}, \citenamefont {Mejia-Diaz}, \citenamefont {Bietenholz},
  \citenamefont {Wiese},\ and\ \citenamefont {Zoller}}]{Laflamme:2015wma}%
  \BibitemOpen
  \bibfield  {author} {\bibinfo {author} {\bibfnamefont {C.}~\bibnamefont
  {Laflamme}}, \bibinfo {author} {\bibfnamefont {W.}~\bibnamefont {Evans}},
  \bibinfo {author} {\bibfnamefont {M.}~\bibnamefont {Dalmonte}}, \bibinfo
  {author} {\bibfnamefont {U.}~\bibnamefont {Gerber}}, \bibinfo {author}
  {\bibfnamefont {H.}~\bibnamefont {Mejia-Diaz}}, \bibinfo {author}
  {\bibfnamefont {W.}~\bibnamefont {Bietenholz}}, \bibinfo {author}
  {\bibfnamefont {U.~J.}\ \bibnamefont {Wiese}}, \ and\ \bibinfo {author}
  {\bibfnamefont {P.}~\bibnamefont {Zoller}},\ }\href {\doibase
  https://doi.org/10.1016/j.aop.2016.03.012} {\bibfield  {journal} {\bibinfo
  {journal} {Annals Phys.}\ }\textbf {\bibinfo {volume} {370}},\ \bibinfo
  {pages} {117} (\bibinfo {year} {2016})},\ \Eprint
  {http://arxiv.org/abs/1507.06788} {arXiv:1507.06788 [quant-ph]} \BibitemShut
  {NoStop}%
\bibitem [{\citenamefont {{Garaud}}\ \emph {et~al.}(2011)\citenamefont
  {{Garaud}}, \citenamefont {{Carlstr{\"o}m}},\ and\ \citenamefont
  {{Babaev}}}]{garaud2011topological}%
  \BibitemOpen
  \bibfield  {author} {\bibinfo {author} {\bibfnamefont {J.}~\bibnamefont
  {{Garaud}}}, \bibinfo {author} {\bibfnamefont {J.}~\bibnamefont
  {{Carlstr{\"o}m}}}, \ and\ \bibinfo {author} {\bibfnamefont {E.}~\bibnamefont
  {{Babaev}}},\ }\href {\doibase 10.1103/PhysRevLett.107.197001} {\bibfield
  {journal} {\bibinfo  {journal} {\prl}\ }\textbf {\bibinfo {volume} {107}},\
  \bibinfo {eid} {197001} (\bibinfo {year} {2011})},\ \Eprint
  {http://arxiv.org/abs/1107.0995} {arXiv:1107.0995 [cond-mat.supr-con]}
  \BibitemShut {NoStop}%
\bibitem [{\citenamefont {Garaud}\ \emph {et~al.}(2013)\citenamefont {Garaud},
  \citenamefont {Carlstr\"om}, \citenamefont {Babaev},\ and\ \citenamefont
  {Speight}}]{Garaud:2012pn}%
  \BibitemOpen
  \bibfield  {author} {\bibinfo {author} {\bibfnamefont {J.}~\bibnamefont
  {Garaud}}, \bibinfo {author} {\bibfnamefont {J.}~\bibnamefont {Carlstr\"om}},
  \bibinfo {author} {\bibfnamefont {E.}~\bibnamefont {Babaev}}, \ and\ \bibinfo
  {author} {\bibfnamefont {M.}~\bibnamefont {Speight}},\ }\href
  {http://dx.doi.org/10.1103/PhysRevB.87.014507} {\bibfield  {journal}
  {\bibinfo  {journal} {Phys. Rev. B}\ }\textbf {\bibinfo {volume} {87}},\
  \bibinfo {pages} {014507} (\bibinfo {year} {2013})},\ \Eprint
  {http://arxiv.org/abs/1211.4342} {arXiv:1211.4342} \BibitemShut {NoStop}%
\bibitem [{\citenamefont {{Benfenati}}\ \emph {et~al.}(2022)\citenamefont
  {{Benfenati}}, \citenamefont {{Barkman}},\ and\ \citenamefont
  {{Babaev}}}]{benfenati2022}%
  \BibitemOpen
  \bibfield  {author} {\bibinfo {author} {\bibfnamefont {A.}~\bibnamefont
  {{Benfenati}}}, \bibinfo {author} {\bibfnamefont {M.}~\bibnamefont
  {{Barkman}}}, \ and\ \bibinfo {author} {\bibfnamefont {E.}~\bibnamefont
  {{Babaev}}},\ }\href@noop {} {\  (\bibinfo {year} {2022})},\ \Eprint
  {http://arxiv.org/abs/2204.05242} {arXiv:2204.05242 [cond-mat.supr-con]}
  \BibitemShut {NoStop}%
\bibitem [{\citenamefont {Papanicolaou}(1988)}]{Papanicolaou1988}%
  \BibitemOpen
  \bibfield  {author} {\bibinfo {author} {\bibfnamefont {N.}~\bibnamefont
  {Papanicolaou}},\ }\href {\doibase
  https://doi.org/10.1016/0550-3213(88)90073-9} {\bibfield  {journal} {\bibinfo
   {journal} {Nucl. Phys. B}\ }\textbf {\bibinfo {volume} {305}},\ \bibinfo
  {pages} {367} (\bibinfo {year} {1988})}\BibitemShut {NoStop}%
\bibitem [{\citenamefont {Batista}\ and\ \citenamefont
  {Ortiz}(2004)}]{Batista2004}%
  \BibitemOpen
  \bibfield  {author} {\bibinfo {author} {\bibfnamefont {C.~D.}\ \bibnamefont
  {Batista}}\ and\ \bibinfo {author} {\bibfnamefont {G.}~\bibnamefont
  {Ortiz}},\ }\href {\doibase 10.1080/00018730310001642086} {\bibfield
  {journal} {\bibinfo  {journal} {Adv. Phys.}\ }\textbf {\bibinfo {volume}
  {53}},\ \bibinfo {pages} {1} (\bibinfo {year} {2004})}\BibitemShut {NoStop}%
\bibitem [{\citenamefont {Tsunetsugu}\ and\ \citenamefont
  {Arikawa}(2006)}]{Tsunetsugu2006}%
  \BibitemOpen
  \bibfield  {author} {\bibinfo {author} {\bibfnamefont {H.}~\bibnamefont
  {Tsunetsugu}}\ and\ \bibinfo {author} {\bibfnamefont {M.}~\bibnamefont
  {Arikawa}},\ }\href {\doibase 10.1143/JPSJ.75.083701} {\bibfield  {journal}
  {\bibinfo  {journal} {J. Phys. Soc. Jpn.}\ }\textbf {\bibinfo {volume}
  {75}},\ \bibinfo {pages} {083701} (\bibinfo {year} {2006})}\BibitemShut
  {NoStop}%
\bibitem [{\citenamefont {{A. L\"auchli, F. Mila, and K.
  Penc}}(2006)}]{Lauchli2006}%
  \BibitemOpen
  \bibfield  {author} {\bibinfo {author} {\bibnamefont {{A. L\"auchli, F. Mila,
  and K. Penc}}},\ }\href {\doibase 10.1103/PhysRevLett.97.087205} {\bibfield
  {journal} {\bibinfo  {journal} {Phys. Rev. Lett.}\ }\textbf {\bibinfo
  {volume} {97}},\ \bibinfo {pages} {087205} (\bibinfo {year}
  {2006})}\BibitemShut {NoStop}%
\bibitem [{\citenamefont {{T. T\'oth, A. M. L\"auchli, F. Mila, and K.
  Penc}}(2010)}]{Toth2010}%
  \BibitemOpen
  \bibfield  {author} {\bibinfo {author} {\bibnamefont {{T. T\'oth, A. M.
  L\"auchli, F. Mila, and K. Penc}}},\ }\href {\doibase
  https://doi.org/10.1103/PhysRevLett.105.265301} {\bibfield  {journal}
  {\bibinfo  {journal} {Phys. Rev. Lett.}\ }\textbf {\bibinfo {volume} {105}},\
  \bibinfo {pages} {265301} (\bibinfo {year} {2010})}\BibitemShut {NoStop}%
\bibitem [{\citenamefont {{Penc}}\ and\ \citenamefont
  {{L{\"a}uchli}}(2011)}]{penc2011spin}%
  \BibitemOpen
  \bibfield  {author} {\bibinfo {author} {\bibfnamefont {K.}~\bibnamefont
  {{Penc}}}\ and\ \bibinfo {author} {\bibfnamefont {A.~M.}\ \bibnamefont
  {{L{\"a}uchli}}},\ }in\ \href {\doibase 10.1007/978-3-642-10589-0\_13} {\emph
  {\bibinfo {booktitle} {Introduction to Frustrated Magnetism: Materials,
  Experiments, Theory}}},\ Vol.\ \bibinfo {volume} {164},\ \bibinfo {editor}
  {edited by\ \bibinfo {editor} {\bibfnamefont {C.}~\bibnamefont {{Lacroix}}},
  \bibinfo {editor} {\bibfnamefont {P.}~\bibnamefont {{Mendels}}}, \ and\
  \bibinfo {editor} {\bibfnamefont {F.}~\bibnamefont {{Mila}}}}\ (\bibinfo
  {year} {2011})\ p.\ \bibinfo {pages} {331}\BibitemShut {NoStop}%
\bibitem [{\citenamefont {{B. Bauer, P. Corboz, A. L\"auchli, L. Messio, K.
  Penc, M. Troyer, and F. Mila}}(2012)}]{Bauer2012}%
  \BibitemOpen
  \bibfield  {author} {\bibinfo {author} {\bibnamefont {{B. Bauer, P. Corboz,
  A. L\"auchli, L. Messio, K. Penc, M. Troyer, and F. Mila}}},\ }\href
  {\doibase 10.1103/PhysRevB.85.125116} {\bibfield  {journal} {\bibinfo
  {journal} {Phys. Rev. B}\ }\textbf {\bibinfo {volume} {85}},\ \bibinfo
  {pages} {125116} (\bibinfo {year} {2012})}\BibitemShut {NoStop}%
\bibitem [{\citenamefont {Ivanov}\ and\ \citenamefont
  {Kolezhuk}(2003)}]{Ivanov2003}%
  \BibitemOpen
  \bibfield  {author} {\bibinfo {author} {\bibfnamefont {B.~A.}\ \bibnamefont
  {Ivanov}}\ and\ \bibinfo {author} {\bibfnamefont {A.~K.}\ \bibnamefont
  {Kolezhuk}},\ }\href {\doibase 10.1103/PhysRevB.68.052401} {\bibfield
  {journal} {\bibinfo  {journal} {Phys. Rev. B}\ }\textbf {\bibinfo {volume}
  {68}},\ \bibinfo {pages} {052401} (\bibinfo {year} {2003})}\BibitemShut
  {NoStop}%
\bibitem [{\citenamefont {Ivanov}\ and\ \citenamefont
  {Khymyn}(2007)}]{Ivanov2007}%
  \BibitemOpen
  \bibfield  {author} {\bibinfo {author} {\bibfnamefont {B.~A.}\ \bibnamefont
  {Ivanov}}\ and\ \bibinfo {author} {\bibfnamefont {R.~S.}\ \bibnamefont
  {Khymyn}},\ }\href {\doibase 10.1134/S106377610702015X} {\bibfield  {journal}
  {\bibinfo  {journal} {J. Exp. Theor. Phys.}\ }\textbf {\bibinfo {volume}
  {104}},\ \bibinfo {pages} {307} (\bibinfo {year} {2007})}\BibitemShut
  {NoStop}%
\bibitem [{\citenamefont {Ivanov}\ \emph {et~al.}(2008)\citenamefont {Ivanov},
  \citenamefont {Khymyn},\ and\ \citenamefont {Kolezhuk}}]{ivanov2008pairing}%
  \BibitemOpen
  \bibfield  {author} {\bibinfo {author} {\bibfnamefont {B.~A.}\ \bibnamefont
  {Ivanov}}, \bibinfo {author} {\bibfnamefont {R.~S.}\ \bibnamefont {Khymyn}},
  \ and\ \bibinfo {author} {\bibfnamefont {A.~K.}\ \bibnamefont {Kolezhuk}},\
  }\href {\doibase 10.1103/PhysRevLett.100.047203} {\bibfield  {journal}
  {\bibinfo  {journal} {Phys. Rev. Lett.}\ }\textbf {\bibinfo {volume} {100}},\
  \bibinfo {pages} {047203} (\bibinfo {year} {2008})},\ \Eprint
  {http://arxiv.org/abs/0710.4557} {arXiv:0710.4557 [cond-mat.str-el]}
  \BibitemShut {NoStop}%
\bibitem [{\citenamefont {Galkina}\ \emph {et~al.}(2015)\citenamefont
  {Galkina}, \citenamefont {Ivanov}, \citenamefont {Kosmachev},\ and\
  \citenamefont {Fridman}}]{Galkina2015}%
  \BibitemOpen
  \bibfield  {author} {\bibinfo {author} {\bibfnamefont {E.~G.}\ \bibnamefont
  {Galkina}}, \bibinfo {author} {\bibfnamefont {B.~A.}\ \bibnamefont {Ivanov}},
  \bibinfo {author} {\bibfnamefont {O.~A.}\ \bibnamefont {Kosmachev}}, \ and\
  \bibinfo {author} {\bibfnamefont {Y.~A.}\ \bibnamefont {Fridman}},\ }\href
  {\doibase 10.1063/1.4921470} {\bibfield  {journal} {\bibinfo  {journal} {Low
  Temp. Phys.}\ }\textbf {\bibinfo {volume} {41}},\ \bibinfo {pages} {382}
  (\bibinfo {year} {2015})}\BibitemShut {NoStop}%
\bibitem [{\citenamefont {Ueda}\ \emph {et~al.}(2016)\citenamefont {Ueda},
  \citenamefont {Akagi},\ and\ \citenamefont {Shannon}}]{Ueda_2016}%
  \BibitemOpen
  \bibfield  {author} {\bibinfo {author} {\bibfnamefont {H.~T.}\ \bibnamefont
  {Ueda}}, \bibinfo {author} {\bibfnamefont {Y.}~\bibnamefont {Akagi}}, \ and\
  \bibinfo {author} {\bibfnamefont {N.}~\bibnamefont {Shannon}},\ }\href
  {http://dx.doi.org/10.1103/PhysRevA.93.021606} {\bibfield  {journal}
  {\bibinfo  {journal} {Phys. Rev. A}\ }\textbf {\bibinfo {volume} {93}}
  (\bibinfo {year} {2016})},\ \Eprint {http://arxiv.org/abs/1511.06515}
  {arXiv:1511.06515} \BibitemShut {NoStop}%
\bibitem [{\citenamefont {Takano}\ and\ \citenamefont
  {Tsunetsugu}(2011)}]{Takano2011}%
  \BibitemOpen
  \bibfield  {author} {\bibinfo {author} {\bibfnamefont {J.}~\bibnamefont
  {Takano}}\ and\ \bibinfo {author} {\bibfnamefont {H.}~\bibnamefont
  {Tsunetsugu}},\ }\href {\doibase 10.1143/JPSJ.80.094707} {\bibfield
  {journal} {\bibinfo  {journal} {J. Phys. Soc. Jpn.}\ }\textbf {\bibinfo
  {volume} {80}},\ \bibinfo {pages} {094707} (\bibinfo {year}
  {2011})}\BibitemShut {NoStop}%
\bibitem [{\citenamefont {Grover}\ and\ \citenamefont
  {Senthil}(2011)}]{Grover2011}%
  \BibitemOpen
  \bibfield  {author} {\bibinfo {author} {\bibfnamefont {T.}~\bibnamefont
  {Grover}}\ and\ \bibinfo {author} {\bibfnamefont {T.}~\bibnamefont
  {Senthil}},\ }\href {\doibase 10.1103/PhysRevLett.107.077203} {\bibfield
  {journal} {\bibinfo  {journal} {Phys. Rev. Lett.}\ }\textbf {\bibinfo
  {volume} {107}},\ \bibinfo {pages} {077203} (\bibinfo {year}
  {2011})}\BibitemShut {NoStop}%
\bibitem [{\citenamefont {Xu}\ and\ \citenamefont {Ludwig}(2012)}]{Xu2012}%
  \BibitemOpen
  \bibfield  {author} {\bibinfo {author} {\bibfnamefont {C.}~\bibnamefont
  {Xu}}\ and\ \bibinfo {author} {\bibfnamefont {A.~W.~W.}\ \bibnamefont
  {Ludwig}},\ }\href {\doibase 10.1103/PhysRevLett.108.047202} {\bibfield
  {journal} {\bibinfo  {journal} {Phys. Rev. Lett.}\ }\textbf {\bibinfo
  {volume} {108}},\ \bibinfo {pages} {047202} (\bibinfo {year}
  {2012})}\BibitemShut {NoStop}%
\bibitem [{\citenamefont {Hu}\ \emph {et~al.}(2014)\citenamefont {Hu},
  \citenamefont {Turner}, \citenamefont {Penc},\ and\ \citenamefont
  {Pollmann}}]{Hu2014}%
  \BibitemOpen
  \bibfield  {author} {\bibinfo {author} {\bibfnamefont {S.}~\bibnamefont
  {Hu}}, \bibinfo {author} {\bibfnamefont {A.~M.}\ \bibnamefont {Turner}},
  \bibinfo {author} {\bibfnamefont {K.}~\bibnamefont {Penc}}, \ and\ \bibinfo
  {author} {\bibfnamefont {F.}~\bibnamefont {Pollmann}},\ }\href {\doibase
  https://doi.org/10.1103/PhysRevLett.113.027202} {\bibfield  {journal}
  {\bibinfo  {journal} {Phys. Rev. Lett.}\ }\textbf {\bibinfo {volume} {113}},\
  \bibinfo {pages} {027202} (\bibinfo {year} {2014})}\BibitemShut {NoStop}%
\bibitem [{\citenamefont {{Remund}}\ \emph {et~al.}(2022)\citenamefont
  {{Remund}}, \citenamefont {{Pohle}}, \citenamefont {{Akagi}}, \citenamefont
  {{Romh{\'a}nyi}},\ and\ \citenamefont {{Shannon}}}]{Remund2022}%
  \BibitemOpen
  \bibfield  {author} {\bibinfo {author} {\bibfnamefont {K.}~\bibnamefont
  {{Remund}}}, \bibinfo {author} {\bibfnamefont {R.}~\bibnamefont {{Pohle}}},
  \bibinfo {author} {\bibfnamefont {Y.}~\bibnamefont {{Akagi}}}, \bibinfo
  {author} {\bibfnamefont {J.}~\bibnamefont {{Romh{\'a}nyi}}}, \ and\ \bibinfo
  {author} {\bibfnamefont {N.}~\bibnamefont {{Shannon}}},\ }\href {\doibase
  10.1103/PhysRevResearch.4.033106} {\bibfield  {journal} {\bibinfo  {journal}
  {Phys. Rev. Research}\ }\textbf {\bibinfo {volume} {4}},\ \bibinfo {eid}
  {033106} (\bibinfo {year} {2022})}\BibitemShut {NoStop}%
\bibitem [{\citenamefont {Imambekov}\ \emph {et~al.}(2003)\citenamefont
  {Imambekov}, \citenamefont {Lukin},\ and\ \citenamefont
  {Demler}}]{imambekov2003spin}%
  \BibitemOpen
  \bibfield  {author} {\bibinfo {author} {\bibfnamefont {A.}~\bibnamefont
  {Imambekov}}, \bibinfo {author} {\bibfnamefont {M.}~\bibnamefont {Lukin}}, \
  and\ \bibinfo {author} {\bibfnamefont {E.}~\bibnamefont {Demler}},\ }\href
  {\doibase 10.1103/PhysRevA.68.063602} {\bibfield  {journal} {\bibinfo
  {journal} {Phys. Rev. A}\ }\textbf {\bibinfo {volume} {68}},\ \bibinfo
  {pages} {063602} (\bibinfo {year} {2003})},\ \Eprint
  {http://arxiv.org/abs/cond-mat/0306204} {arXiv:cond-mat/0306204
  [cond-mat.soft]} \BibitemShut {NoStop}%
\bibitem [{\citenamefont {Juzeli\ifmmode~\bar{u}\else \={u}\fi{}nas}\ \emph
  {et~al.}(2010)\citenamefont {Juzeli\ifmmode~\bar{u}\else \={u}\fi{}nas},
  \citenamefont {Ruseckas},\ and\ \citenamefont
  {Dalibard}}]{PhysRevA.81.053403}%
  \BibitemOpen
  \bibfield  {author} {\bibinfo {author} {\bibfnamefont {G.}~\bibnamefont
  {Juzeli\ifmmode~\bar{u}\else \={u}\fi{}nas}}, \bibinfo {author}
  {\bibfnamefont {J.}~\bibnamefont {Ruseckas}}, \ and\ \bibinfo {author}
  {\bibfnamefont {J.}~\bibnamefont {Dalibard}},\ }\href {\doibase
  https://doi.org/10.1103/PhysRevA.81.053403} {\bibfield  {journal} {\bibinfo
  {journal} {Phys. Rev. A}\ }\textbf {\bibinfo {volume} {81}},\ \bibinfo
  {pages} {053403} (\bibinfo {year} {2010})}\BibitemShut {NoStop}%
\bibitem [{\citenamefont {Dalibard}\ \emph {et~al.}(2011)\citenamefont
  {Dalibard}, \citenamefont {Gerbier}, \citenamefont {Juzeliunas},\ and\
  \citenamefont {Ohberg}}]{Dalibard:2010ph}%
  \BibitemOpen
  \bibfield  {author} {\bibinfo {author} {\bibfnamefont {J.}~\bibnamefont
  {Dalibard}}, \bibinfo {author} {\bibfnamefont {F.}~\bibnamefont {Gerbier}},
  \bibinfo {author} {\bibfnamefont {G.}~\bibnamefont {Juzeliunas}}, \ and\
  \bibinfo {author} {\bibfnamefont {P.}~\bibnamefont {Ohberg}},\ }\href
  {\doibase 10.1103/RevModPhys.83.1523} {\bibfield  {journal} {\bibinfo
  {journal} {Rev. Mod. Phys.}\ }\textbf {\bibinfo {volume} {83}},\ \bibinfo
  {pages} {1523} (\bibinfo {year} {2011})},\ \Eprint
  {http://arxiv.org/abs/1008.5378} {arXiv:1008.5378 [cond-mat.quant-gas]}
  \BibitemShut {NoStop}%
\bibitem [{\citenamefont {Goldman}\ \emph {et~al.}(2014)\citenamefont
  {Goldman}, \citenamefont {Juzeli\={u}nas}, \citenamefont {\"Ohberg},\ and\
  \citenamefont {Spielman}}]{Goldman:2013xka}%
  \BibitemOpen
  \bibfield  {author} {\bibinfo {author} {\bibfnamefont {N.}~\bibnamefont
  {Goldman}}, \bibinfo {author} {\bibfnamefont {G.}~\bibnamefont
  {Juzeli\={u}nas}}, \bibinfo {author} {\bibfnamefont {P.}~\bibnamefont
  {\"Ohberg}}, \ and\ \bibinfo {author} {\bibfnamefont {I.~B.}\ \bibnamefont
  {Spielman}},\ }\href {\doibase 10.1088/0034-4885/77/12/126401} {\bibfield
  {journal} {\bibinfo  {journal} {Rept. Prog. Phys.}\ }\textbf {\bibinfo
  {volume} {77}},\ \bibinfo {pages} {126401} (\bibinfo {year} {2014})},\
  \Eprint {http://arxiv.org/abs/1308.6533} {arXiv:1308.6533
  [cond-mat.quant-gas]} \BibitemShut {NoStop}%
\bibitem [{\citenamefont {Zhai}(2015)}]{Zhai:2014gna}%
  \BibitemOpen
  \bibfield  {author} {\bibinfo {author} {\bibfnamefont {H.}~\bibnamefont
  {Zhai}},\ }\href {\doibase 10.1088/0034-4885/78/2/026001} {\bibfield
  {journal} {\bibinfo  {journal} {Rept. Prog. Phys.}\ }\textbf {\bibinfo
  {volume} {78}},\ \bibinfo {pages} {026001} (\bibinfo {year} {2015})},\
  \Eprint {http://arxiv.org/abs/1403.8021} {arXiv:1403.8021
  [cond-mat.quant-gas]} \BibitemShut {NoStop}%
\bibitem [{Note1()}]{Note1}%
  \BibitemOpen
  \bibinfo {note} {Stabilization of three-dimensional Skyrmions in ultracold
  atomic gases with an SOC was proposed in Ref.~\cite
  {Kawakami:2012zw}.}\BibitemShut {Stop}%
\bibitem [{\citenamefont {Klebanov}(1985)}]{Klebanov:1985qi}%
  \BibitemOpen
  \bibfield  {author} {\bibinfo {author} {\bibfnamefont {I.~R.}\ \bibnamefont
  {Klebanov}},\ }\href {\doibase 10.1016/0550-3213(85)90068-9} {\bibfield
  {journal} {\bibinfo  {journal} {Nucl. Phys. B}\ }\textbf {\bibinfo {volume}
  {262}},\ \bibinfo {pages} {133} (\bibinfo {year} {1985})}\BibitemShut
  {NoStop}%
\bibitem [{\citenamefont {Kugler}\ and\ \citenamefont
  {Shtrikman}(1989)}]{Kugler:1989uc}%
  \BibitemOpen
  \bibfield  {author} {\bibinfo {author} {\bibfnamefont {M.}~\bibnamefont
  {Kugler}}\ and\ \bibinfo {author} {\bibfnamefont {S.}~\bibnamefont
  {Shtrikman}},\ }\href {\doibase 10.1103/PhysRevD.40.3421} {\bibfield
  {journal} {\bibinfo  {journal} {Phys. Rev. D}\ }\textbf {\bibinfo {volume}
  {40}},\ \bibinfo {pages} {3421} (\bibinfo {year} {1989})}\BibitemShut
  {NoStop}%
\bibitem [{\citenamefont {Rho}\ and\ \citenamefont
  {Zahed}(2016{\natexlab{b}})}]{parkvento}%
  \BibitemOpen
  \bibinfo {editor} {\bibfnamefont {M.}~\bibnamefont {Rho}}\ and\ \bibinfo
  {editor} {\bibfnamefont {I.}~\bibnamefont {Zahed}},\ eds.,\ \enquote
  {\bibinfo {title} {{Skyrmion approach to finite density and temperature}},}\
  in\ \href@noop {} {\emph {\bibinfo {booktitle} {{The Multifaceted
  Skyrmions}}}}\ (\bibinfo  {publisher} {World Scientific},\ \bibinfo {year}
  {2016})\ Chap.~\bibinfo {chapter} {7}, pp.\ \bibinfo {pages} {131--162},\
  \bibinfo {edition} {2nd}\ ed.\BibitemShut {Stop}%
\bibitem [{\citenamefont {Akagi}\ \emph
  {et~al.}(2021{\natexlab{a}})\citenamefont {Akagi}, \citenamefont {Amari},
  \citenamefont {Sawado},\ and\ \citenamefont {Shnir}}]{Akagi:2021dpk}%
  \BibitemOpen
  \bibfield  {author} {\bibinfo {author} {\bibfnamefont {Y.}~\bibnamefont
  {Akagi}}, \bibinfo {author} {\bibfnamefont {Y.}~\bibnamefont {Amari}},
  \bibinfo {author} {\bibfnamefont {N.}~\bibnamefont {Sawado}}, \ and\ \bibinfo
  {author} {\bibfnamefont {Y.}~\bibnamefont {Shnir}},\ }\href
  {http://dx.doi.org/10.1103/PhysRevD.103.065008} {\bibfield  {journal}
  {\bibinfo  {journal} {Phys. Rev. D}\ }\textbf {\bibinfo {volume} {103}},\
  \bibinfo {pages} {065008} (\bibinfo {year} {2021}{\natexlab{a}})},\ \Eprint
  {http://arxiv.org/abs/2101.10566} {arXiv:2101.10566} \BibitemShut {NoStop}%
\bibitem [{Note2()}]{Note2}%
  \BibitemOpen
  \bibinfo {note} {It may still be an experimental challenge to control the DM
  term in two (spatial) dimensions.}\BibitemShut {Stop}%
\bibitem [{\citenamefont {Gleiser}\ and\ \citenamefont
  {Stamatopoulos}(2012)}]{Gleiser:2011di}%
  \BibitemOpen
  \bibfield  {author} {\bibinfo {author} {\bibfnamefont {M.}~\bibnamefont
  {Gleiser}}\ and\ \bibinfo {author} {\bibfnamefont {N.}~\bibnamefont
  {Stamatopoulos}},\ }\href {\doibase 10.1016/j.physletb.2012.05.064}
  {\bibfield  {journal} {\bibinfo  {journal} {Phys. Lett. B}\ }\textbf
  {\bibinfo {volume} {713}},\ \bibinfo {pages} {304} (\bibinfo {year}
  {2012})},\ \Eprint {http://arxiv.org/abs/1111.5597} {arXiv:1111.5597
  [hep-th]} \BibitemShut {NoStop}%
\bibitem [{\citenamefont {Gleiser}\ and\ \citenamefont
  {Sowinski}(2015)}]{Gleiser:2015aav}%
  \BibitemOpen
  \bibfield  {author} {\bibinfo {author} {\bibfnamefont {M.}~\bibnamefont
  {Gleiser}}\ and\ \bibinfo {author} {\bibfnamefont {D.}~\bibnamefont
  {Sowinski}},\ }\href {\doibase 10.1016/j.physletb.2015.05.058} {\bibfield
  {journal} {\bibinfo  {journal} {Phys. Lett. B}\ }\textbf {\bibinfo {volume}
  {747}},\ \bibinfo {pages} {125} (\bibinfo {year} {2015})},\ \Eprint
  {http://arxiv.org/abs/1501.06800} {arXiv:1501.06800 [cond-mat.stat-mech]}
  \BibitemShut {NoStop}%
\bibitem [{\citenamefont {Bazeia}\ \emph {et~al.}(2019)\citenamefont {Bazeia},
  \citenamefont {Moreira},\ and\ \citenamefont
  {Rodrigues}}]{bazeia2019configurational}%
  \BibitemOpen
  \bibfield  {author} {\bibinfo {author} {\bibfnamefont {D.}~\bibnamefont
  {Bazeia}}, \bibinfo {author} {\bibfnamefont {D.}~\bibnamefont {Moreira}}, \
  and\ \bibinfo {author} {\bibfnamefont {E.}~\bibnamefont {Rodrigues}},\ }\href
  {\doibase 10.1016/j.jmmm.2018.12.033} {\bibfield  {journal} {\bibinfo
  {journal} {J. Magn. Magn. Mater}\ }\textbf {\bibinfo {volume} {475}},\
  \bibinfo {pages} {734} (\bibinfo {year} {2019})}\BibitemShut {NoStop}%
\bibitem [{\citenamefont {Bazeia}\ and\ \citenamefont
  {Rodrigues}(2021)}]{bazeia2021configurational}%
  \BibitemOpen
  \bibfield  {author} {\bibinfo {author} {\bibfnamefont {D.}~\bibnamefont
  {Bazeia}}\ and\ \bibinfo {author} {\bibfnamefont {E.}~\bibnamefont
  {Rodrigues}},\ }\href {\doibase
  https://doi.org/10.1016/j.physleta.2021.127170} {\bibfield  {journal}
  {\bibinfo  {journal} {Phys. Lett. A}\ }\textbf {\bibinfo {volume} {392}},\
  \bibinfo {pages} {127170} (\bibinfo {year} {2021})}\BibitemShut {NoStop}%
\bibitem [{\citenamefont {{Kobayashi}}\ \emph {et~al.}(2014)\citenamefont
  {{Kobayashi}}, \citenamefont {{Nakano}},\ and\ \citenamefont
  {{Nitta}}}]{kobayashi2014color}%
  \BibitemOpen
  \bibfield  {author} {\bibinfo {author} {\bibfnamefont {M.}~\bibnamefont
  {{Kobayashi}}}, \bibinfo {author} {\bibfnamefont {E.}~\bibnamefont
  {{Nakano}}}, \ and\ \bibinfo {author} {\bibfnamefont {M.}~\bibnamefont
  {{Nitta}}},\ }\href {\doibase 10.1007/JHEP06(2014)130} {\bibfield  {journal}
  {\bibinfo  {journal} {JHEP}\ }\textbf {\bibinfo {volume} {2014}},\ \bibinfo
  {eid} {130} (\bibinfo {year} {2014})},\ \Eprint
  {http://arxiv.org/abs/1311.2399} {arXiv:1311.2399 [hep-ph]} \BibitemShut
  {NoStop}%
\bibitem [{\citenamefont {{Stamper-Kurn}}\ and\ \citenamefont
  {{Ueda}}(2013)}]{stamper2013spinor}%
  \BibitemOpen
  \bibfield  {author} {\bibinfo {author} {\bibfnamefont {D.~M.}\ \bibnamefont
  {{Stamper-Kurn}}}\ and\ \bibinfo {author} {\bibfnamefont {M.}~\bibnamefont
  {{Ueda}}},\ }\href {\doibase 10.1103/RevModPhys.85.1191} {\bibfield
  {journal} {\bibinfo  {journal} {Rev. Mod. Phys}\ }\textbf {\bibinfo {volume}
  {85}},\ \bibinfo {pages} {1191} (\bibinfo {year} {2013})}\BibitemShut
  {NoStop}%
\bibitem [{\citenamefont {Kawaguchi}\ and\ \citenamefont
  {Ueda}(2012)}]{Kawaguchi:2012ii}%
  \BibitemOpen
  \bibfield  {author} {\bibinfo {author} {\bibfnamefont {Y.}~\bibnamefont
  {Kawaguchi}}\ and\ \bibinfo {author} {\bibfnamefont {M.}~\bibnamefont
  {Ueda}},\ }\href {\doibase 10.1016/j.physrep.2012.07.005} {\bibfield
  {journal} {\bibinfo  {journal} {Phys. Rept.}\ }\textbf {\bibinfo {volume}
  {520}},\ \bibinfo {pages} {253} (\bibinfo {year} {2012})}\BibitemShut
  {NoStop}%
\bibitem [{\citenamefont {Akagi}\ \emph
  {et~al.}(2021{\natexlab{b}})\citenamefont {Akagi}, \citenamefont {Amari},
  \citenamefont {Gudnason}, \citenamefont {Nitta},\ and\ \citenamefont
  {Shnir}}]{Akagi:2021lva}%
  \BibitemOpen
  \bibfield  {author} {\bibinfo {author} {\bibfnamefont {Y.}~\bibnamefont
  {Akagi}}, \bibinfo {author} {\bibfnamefont {Y.}~\bibnamefont {Amari}},
  \bibinfo {author} {\bibfnamefont {S.~B.}\ \bibnamefont {Gudnason}}, \bibinfo
  {author} {\bibfnamefont {M.}~\bibnamefont {Nitta}}, \ and\ \bibinfo {author}
  {\bibfnamefont {Y.}~\bibnamefont {Shnir}},\ }\href {\doibase
  https://doi.org/10.1007/JHEP11(2021)194} {\bibfield  {journal} {\bibinfo
  {journal} {JHEP}\ }\textbf {\bibinfo {volume} {11}},\ \bibinfo {pages} {194}
  (\bibinfo {year} {2021}{\natexlab{b}})},\ \Eprint
  {http://arxiv.org/abs/2107.13777} {arXiv:2107.13777 [hep-th]} \BibitemShut
  {NoStop}%
\bibitem [{\citenamefont {Graß}\ \emph {et~al.}(2014)\citenamefont {Graß},
  \citenamefont {Chhajlany}, \citenamefont {Muschik},\ and\ \citenamefont
  {Lewenstein}}]{Gra__2014}%
  \BibitemOpen
  \bibfield  {author} {\bibinfo {author} {\bibfnamefont {T.}~\bibnamefont
  {Graß}}, \bibinfo {author} {\bibfnamefont {R.~W.}\ \bibnamefont
  {Chhajlany}}, \bibinfo {author} {\bibfnamefont {C.~A.}\ \bibnamefont
  {Muschik}}, \ and\ \bibinfo {author} {\bibfnamefont {M.}~\bibnamefont
  {Lewenstein}},\ }\href
  {https://journals.aps.org/prb/abstract/10.1103/PhysRevB.90.195127} {\bibfield
   {journal} {\bibinfo  {journal} {Phys. Rev. B}\ }\textbf {\bibinfo {volume}
  {90}},\ \bibinfo {pages} {195127} (\bibinfo {year} {2014})},\ \Eprint
  {http://arxiv.org/abs/1408.0769} {arXiv:1408.0769} \BibitemShut {NoStop}%
\bibitem [{\citenamefont {Stoudenmire}\ \emph {et~al.}(2009)\citenamefont
  {Stoudenmire}, \citenamefont {Trebst},\ and\ \citenamefont
  {Balents}}]{Stoudenmire2009}%
  \BibitemOpen
  \bibfield  {author} {\bibinfo {author} {\bibfnamefont {E.~M.}\ \bibnamefont
  {Stoudenmire}}, \bibinfo {author} {\bibfnamefont {S.}~\bibnamefont {Trebst}},
  \ and\ \bibinfo {author} {\bibfnamefont {L.}~\bibnamefont {Balents}},\ }\href
  {\doibase 10.1103/PhysRevB.79.214436} {\bibfield  {journal} {\bibinfo
  {journal} {Phys. Rev. B}\ }\textbf {\bibinfo {volume} {79}},\ \bibinfo
  {pages} {214436} (\bibinfo {year} {2009})}\BibitemShut {NoStop}%
\bibitem [{\citenamefont {Yamamoto}\ \emph {et~al.}(2020)\citenamefont
  {Yamamoto}, \citenamefont {Suzuki}, \citenamefont {Marmorini}, \citenamefont
  {Okazaki},\ and\ \citenamefont {Furukawa}}]{Yamamoto2020}%
  \BibitemOpen
  \bibfield  {author} {\bibinfo {author} {\bibfnamefont {D.}~\bibnamefont
  {Yamamoto}}, \bibinfo {author} {\bibfnamefont {C.}~\bibnamefont {Suzuki}},
  \bibinfo {author} {\bibfnamefont {G.}~\bibnamefont {Marmorini}}, \bibinfo
  {author} {\bibfnamefont {S.}~\bibnamefont {Okazaki}}, \ and\ \bibinfo
  {author} {\bibfnamefont {N.}~\bibnamefont {Furukawa}},\ }\href {\doibase
  https://doi.org/10.1103/PhysRevLett.125.057204} {\bibfield  {journal}
  {\bibinfo  {journal} {Phys. Rev. Lett.}\ }\textbf {\bibinfo {volume} {125}},\
  \bibinfo {pages} {057204} (\bibinfo {year} {2020})}\BibitemShut {NoStop}%
\bibitem [{\citenamefont {Tanaka}\ and\ \citenamefont
  {Hotta}(2020)}]{Tanaka2020}%
  \BibitemOpen
  \bibfield  {author} {\bibinfo {author} {\bibfnamefont {K.}~\bibnamefont
  {Tanaka}}\ and\ \bibinfo {author} {\bibfnamefont {C.}~\bibnamefont {Hotta}},\
  }\href {\doibase 10.1103/PhysRevB.102.140401} {\bibfield  {journal} {\bibinfo
   {journal} {Phys. Rev. B}\ }\textbf {\bibinfo {volume} {102}},\ \bibinfo
  {pages} {140401} (\bibinfo {year} {2020})}\BibitemShut {NoStop}%
\bibitem [{\citenamefont {Amari}\ and\ \citenamefont
  {Sawado}(2018{\natexlab{a}})}]{Amari:2017qnb}%
  \BibitemOpen
  \bibfield  {author} {\bibinfo {author} {\bibfnamefont {Y.}~\bibnamefont
  {Amari}}\ and\ \bibinfo {author} {\bibfnamefont {N.}~\bibnamefont {Sawado}},\
  }\href {http://dx.doi.org/10.1103/PhysRevD.97.065012} {\bibfield  {journal}
  {\bibinfo  {journal} {Phys. Rev. D}\ }\textbf {\bibinfo {volume} {97}},\
  \bibinfo {pages} {065012} (\bibinfo {year} {2018}{\natexlab{a}})},\ \Eprint
  {http://arxiv.org/abs/1711.00933} {arXiv:1711.00933} \BibitemShut {NoStop}%
\bibitem [{\citenamefont {Amari}\ and\ \citenamefont
  {Sawado}(2018{\natexlab{b}})}]{Amari:2018gbq}%
  \BibitemOpen
  \bibfield  {author} {\bibinfo {author} {\bibfnamefont {Y.}~\bibnamefont
  {Amari}}\ and\ \bibinfo {author} {\bibfnamefont {N.}~\bibnamefont {Sawado}},\
  }\href {https://doi.org/10.1016/j.physletb.2018.08.020} {\bibfield  {journal}
  {\bibinfo  {journal} {Phys. Lett. B}\ }\textbf {\bibinfo {volume} {784}},\
  \bibinfo {pages} {294} (\bibinfo {year} {2018}{\natexlab{b}})},\ \Eprint
  {http://arxiv.org/abs/1805.10008} {arXiv:1805.10008} \BibitemShut {NoStop}%
\bibitem [{\citenamefont {Affleck}\ \emph {et~al.}(2022)\citenamefont
  {Affleck}, \citenamefont {Bykov},\ and\ \citenamefont
  {Wamer}}]{Affleck:2021jls}%
  \BibitemOpen
  \bibfield  {author} {\bibinfo {author} {\bibfnamefont {I.}~\bibnamefont
  {Affleck}}, \bibinfo {author} {\bibfnamefont {D.}~\bibnamefont {Bykov}}, \
  and\ \bibinfo {author} {\bibfnamefont {K.}~\bibnamefont {Wamer}},\ }\href
  {\doibase 10.1016/j.physrep.2021.09.004} {\bibfield  {journal} {\bibinfo
  {journal} {Phys. Rept.}\ }\textbf {\bibinfo {volume} {953}},\ \bibinfo
  {pages} {1} (\bibinfo {year} {2022})},\ \Eprint
  {http://arxiv.org/abs/2101.11638} {arXiv:2101.11638 [hep-th]} \BibitemShut
  {NoStop}%
\bibitem [{\citenamefont {Takahashi}\ and\ \citenamefont
  {Tanizaki}(2021)}]{Takahashi:2021lvg}%
  \BibitemOpen
  \bibfield  {author} {\bibinfo {author} {\bibfnamefont {I.}~\bibnamefont
  {Takahashi}}\ and\ \bibinfo {author} {\bibfnamefont {Y.}~\bibnamefont
  {Tanizaki}},\ }\href {\doibase 10.1103/PhysRevB.104.235152} {\bibfield
  {journal} {\bibinfo  {journal} {Phys. Rev. B}\ }\textbf {\bibinfo {volume}
  {104}},\ \bibinfo {pages} {235152} (\bibinfo {year} {2021})},\ \Eprint
  {http://arxiv.org/abs/2109.10051} {arXiv:2109.10051 [cond-mat.str-el]}
  \BibitemShut {NoStop}%
\bibitem [{\citenamefont {Zhang}\ \emph {et~al.}(2022)\citenamefont {Zhang},
  \citenamefont {Wang}, \citenamefont {Dahlbom}, \citenamefont {Barros},\ and\
  \citenamefont {Batista}}]{Zhang:2022lyz}%
  \BibitemOpen
  \bibfield  {author} {\bibinfo {author} {\bibfnamefont {H.}~\bibnamefont
  {Zhang}}, \bibinfo {author} {\bibfnamefont {Z.}~\bibnamefont {Wang}},
  \bibinfo {author} {\bibfnamefont {D.}~\bibnamefont {Dahlbom}}, \bibinfo
  {author} {\bibfnamefont {K.}~\bibnamefont {Barros}}, \ and\ \bibinfo {author}
  {\bibfnamefont {C.~D.}\ \bibnamefont {Batista}},\ }\href@noop {} {\
  (\bibinfo {year} {2022})},\ \Eprint {http://arxiv.org/abs/2203.15248}
  {arXiv:2203.15248 [cond-mat.str-el]} \BibitemShut {NoStop}%
\bibitem [{\citenamefont {Kawakami}\ \emph {et~al.}(2012)\citenamefont
  {Kawakami}, \citenamefont {Mizushima}, \citenamefont {Nitta},\ and\
  \citenamefont {Machida}}]{Kawakami:2012zw}%
  \BibitemOpen
  \bibfield  {author} {\bibinfo {author} {\bibfnamefont {T.}~\bibnamefont
  {Kawakami}}, \bibinfo {author} {\bibfnamefont {T.}~\bibnamefont {Mizushima}},
  \bibinfo {author} {\bibfnamefont {M.}~\bibnamefont {Nitta}}, \ and\ \bibinfo
  {author} {\bibfnamefont {K.}~\bibnamefont {Machida}},\ }\href {\doibase
  https://doi.org/10.1103/PhysRevLett.109.015301} {\bibfield  {journal}
  {\bibinfo  {journal} {Phys. Rev. Lett.}\ }\textbf {\bibinfo {volume} {109}},\
  \bibinfo {pages} {015301} (\bibinfo {year} {2012})},\ \Eprint
  {http://arxiv.org/abs/1204.3177} {arXiv:1204.3177 [cond-mat.quant-gas]}
  \BibitemShut {NoStop}%
\end{thebibliography}%

\onecolumngrid

\section*{Supplemental material: Derivation of the effective Hamiltonian}
\setcounter{equation}{0}
\renewcommand{\theequation}{S.\arabic{equation}}

	This supplemental material is devoted to the derivation of the effective Hamiltonian~\eqref{eq:H} in the main text (ferromagnetic $\SU(3)$ Heisenberg model with a generalized Dzyaloshinskii-Moriya interaction and Zeeman term) from a spin-orbit coupled spin-$1$ Bose-Hubbard model on a square lattice. Let us begin with the Hamiltonian of the form
	\begin{equation}
		\hH = \hH_t + \hH_U, 
		\label{Hamiltonian_gBH}
	\end{equation}
	with
	\begin{align}
		&\hH_t=-t\sum_{\left\langle i,j \right\rangle } \left[\sum_{\sigma,\rho}\hb^\dagger_{i,\sigma} (\Lambda_{i,j})_{\sigma\rho}\hb_{j,\rho}+{\rm H.c.}\right] ,
		\label{Hamiltonian_hop}
		\\
		&\hH_U=\frac{U}{2}\sum_i \hn_i(\hn_i-1) ,
		\label{Hamiltonian_int}
	\end{align}
	where $t$ is the transfer integral, $U$ is the on-site repulsion, 
	$\hb_{i,\sigma}^\dagger$ ($\hb_{i,\sigma}$) is the creation (annihilation) operator of a boson at site $i$ with spin $\sigma\in \{ 0,\pm 1\}$, and $\hn_i=\sum_\sigma \hb_{i,\sigma}^\dagger\hb_{i,\sigma}$ is the particle number operator at site $i$. In addition, $\Lambda_{i,j}\approx \mathbf{1} + \i A_{i,j}$ denotes the Wilson line for the (classical) $\SU(3)$ gauge potential $A_{i,j}$ defined on the link between sites $i$ and $j$, where the gauge potential satisfies $A_{i,j}^\dagger = A_{i,j}$ and $A_{j,i}=-A_{i,j}$.	
	The term in Eq.~\eqref{Hamiltonian_hop} proportional to the gauge potential describes the $\SU(3)$ SOC.
	We show that the ferromagnetic $\SU(3)$ exchange interaction and the generalized DM interaction are simultaneously obtained in the strong coupling limit $0<t \ll U$. The inclusion of the Zeeman term is rather trivial, and thus we put it aside for now. 
	
	In the following, for simplicity, we consider a case where there exists one boson per site.
	When $t=0$, the Hilbert space of the ground state is 
	\begin{equation}
		\calS= \left\{ \ket{..., n_{i}, ...} | \forall i, n_i=1 \right\}.
	\end{equation} 
	For $t>0$, low-energy states are spanned by 
	the subspace of  the Fock space
	\begin{equation}
		\calD= \left\{ \ket{..., n_{i}, ...} | \forall i, n_i\leq 2 \right\},
	\end{equation} 
	which are all configurations with at most two bosons on each site.
	The hopping term restricted to $\calD$, $H_t|_\calD$, can be split into the following three parts:
	\begin{align}
		&\hH_t|_\calD = H_t^0 +H_t^+ + H_t^-,
		\\
		&\hH_t^0 = -t \sum_{\left\langle i,j \right\rangle } \left[\hdel_{n_i,1}\hb_{i,\sigma}^\dagger(\Lambda_{i,j})_{\sigma\rho} \hb_{j,\rho} \hdel_{n_j,1}+ \hdel_{n_i,2}\hb_{i,\sigma}^\dagger(\Lambda_{i,j})_{\sigma \rho} \hb_{j,\rho} \hdel_{n_j,2} + (i\leftrightarrow j )
		\right],
		\\
		&\hH_t^+ = -t \sum_{\left\langle i,j \right\rangle } \left[
		\hdel_{n_i,2}\hb_{i,\sigma}^\dagger(\Lambda_{i,j})_{\sigma\rho} \hb_{j,\rho} \hdel_{n_j,1} 
		+ (i\leftrightarrow j )
		\right],
		\\
		&\hH_t^- = -t \sum_{\left\langle i,j \right\rangle } \left[
		\hdel_{n_i,1}\hb_{i,\sigma}^\dagger(\Lambda_{i,j})_{\sigma\rho} \hb_{j,\rho} \hdel_{n_j,2} 
		+ (i\leftrightarrow j )
		\right],
	\end{align} 
	where $\hdel_{n_i,\nu} \ket{\phi} = \delta_{n_i,\nu}\ket{\phi}$ for $\hn_i\ket{\phi} = n_i\ket{\phi}$ \citeSM{barthel2009magnetism}.
	The term $\hH_t^0$ describes the hopping process leaving the number of doubly occupied sites unchanged, and $\hH_t^+/\hH_t^-$ does the process increasing/decreasing it by one. 
	
	We expand $\hH|_\calD = \hH_t|_\calD + \hH_U$ in terms of the small parameter $t/U$, and derive an effective Hamiltonian to order $t^2/{U}$ using the Schrieffer-Wolff transformation \citeSM{schrieffer1966relation,barthel2009magnetism,zhu2014spin,pixley2017strong}. 
	Let $P_\calS$ be the projector onto the Hilbert space $\calS$.
	Then, we define the effective Hamiltonian $\hH_{\rm eff}$ via
	\begin{equation}
		\hH_{\rm eff}|_\calS := P_\calS \hH_{\rm eff} P_\calS = P_\calS \hH' P_\calS ,
	\end{equation}
	where 
	\begin{align}
		\hH' &= e^{\heta} \hH|_\calD e^{-\heta}
		\notag\\
		&= \hH|_\calD + [\heta, \hH|_\calD] + \frac{1}{2}[\heta,[\heta, \hH|_\calD]] + \cdots,
		\label{Hamiltonian_SW}
	\end{align}
	with $\heta$ being an anti-Hermitian operator.
	If we choose $\heta$ satisfying
	\begin{equation}
		[\heta, \hH_U] = -(\hH_t^+ + \hH_t^- ),
		\label{condition_eta}
	\end{equation}
	the effective Hamiltonian reduces to the form
	\begin{equation}
		\hH_{\rm eff}|_\calS = \frac{1}{2}
		P_\calS[\heta, \hH_t+  \hH_t^- ]P_\calS + \calO(t^3/U^2 ),
		\label{Hamiltonian_eff_eta}
	\end{equation}
	where we used $P_\calS  \hH_U P_\calS =0$ and $P_\calS H_t^0=H_t^0P_\calS =0$.
	One finds that we can define $\heta$ enjoying Eq.~\eqref{condition_eta} as 
	\begin{equation}
		\heta = -\frac{1}{U} \left(\hH_t^+-\hH_t^- \right).
		\label{explicit_form_eta}
	\end{equation}
	Substituting Eq. \eqref{explicit_form_eta} into Eq. \eqref{Hamiltonian_eff_eta}, we obtain
	\begin{align}
		\hH_{\rm eff}|_\calS &= -\frac{t^2}{U} \sum_{\left\langle i,j \right\rangle }\sum_{\sigma, \rho,\sigma' ,\rho'} P_\calS
		\left\{ (\Lambda_{i,j})_{\sigma\rho} (\Lambda_{j,i})_{\sigma'\rho'} 
		\hb_{i,\sigma}^\dagger \hb_{j,\rho} \hb_{j,\sigma'}^\dagger\hb_{i,\rho'}+ (i\leftrightarrow j)
		\right\} P_\calS
		\notag\\
		&\approx -\frac{t^2}{U} \sum_{\left\langle i,j \right\rangle }P_\calS
		\left[ \sum_{\sigma, \rho} 
		\hb_{i,\sigma}^\dagger \hb_{j,\sigma} \hb_{j,\rho}^\dagger\hb_{i,\rho} 
		+ \i \sum_{\sigma, \rho,\tau} (A_{i,j})_{\sigma\rho} 
		\left\{	\hb_{i,\sigma}^\dagger \hb_{j,\rho} \hb_{j,\tau}^\dagger\hb_{i,\tau}  - 	\hb_{i,\tau}^\dagger \hb_{j,\tau} \hb_{j,\sigma}^\dagger\hb_{i,\rho}  \right\}
		+ (i\leftrightarrow j)
		\right] P_\calS,
	\end{align}
	where we used $P_\calS \hdel_{n_i,2} =\hdel_{n_i,2}P_\calS=0$ and omitted higher order terms in the gauge potential.	
	
	We introduce the $\SU(3)$ spin operators defined as
	\begin{equation}
		\hT_i^\alpha = \sum_{\sigma,\rho}(\lambda_\alpha)_{\sigma\rho}\hb_{i,\sigma}^\dagger\hb_{i,\rho} .
	\end{equation}
	Using the identities $\sum_{\alpha=1}^8(\lambda_\alpha)_{\sigma\rho}(\lambda_\alpha)_{\mu\nu}
	=2\delta_{\sigma\nu}\delta_{\rho\mu}-\frac{2}{3}\delta_{\sigma\rho}\delta_{\mu\nu}$ and 
	$\sum_{\beta,\gamma=1}^8f_{\alpha\beta\gamma}\lambda^\beta_{\sigma\rho}\lambda^\gamma_{\mu\nu} = \i (\delta_{\rho\mu}\lambda^\alpha_{\sigma\nu} -\delta_{\sigma\nu}\lambda^\alpha_{\mu\rho})$, we find
	\begin{align}
		&\sum_{\alpha=1}^8\hT^\alpha_i \hT^\alpha_j 
		= 2\sum_{\sigma,\rho} \hb_{i,\sigma}^\dagger\hb_{j,\sigma}\hb_{j,\rho}^\dagger\hb_{i,\rho} - 2 \hn_i - \frac{2}{3}\hn_i\hn_j,
		\\
		&\sum_{\beta,\gamma=1}^8f_{\alpha\beta\gamma}\hT^\beta_i\hT^\gamma_j
		=\i \sum_{\sigma,\rho, \tau}(\lambda_\alpha)_{\sigma\rho} 
		\left\{	\hb_{i,\sigma}^\dagger \hb_{j,\rho} \hb_{j,\tau}^\dagger\hb_{i,\tau}  - 	\hb_{i,\tau}^\dagger \hb_{j,\tau} \hb_{j,\sigma}^\dagger\hb_{i,\rho}  \right\},
	\end{align}
	for $i\neq j$.
	Therefore, expanding the gauge potential as $A_{i,j}=\sum_{\alpha=1}^8A_{i,j}^\alpha\lambda_\alpha$, we find the effective Hamiltonian can be cast into the form
	\begin{equation}
		\begin{split}
			\hH_{\rm eff}
			&\equiv \frac{J}{2}\sum_{\exv{i,j} } 
			\Bigg[\sum_{\alpha=1}^8\hT^\alpha_i\hT^\alpha_j  + 2 \sum_{\alpha,\beta, \gamma=1}^8A_{i,j}^\alpha f_{\alpha\beta\gamma}\hT^\beta_i\hT^\gamma_j \Bigg]
			+ {\rm const.},
		\end{split}
		\label{Hamiltonian_T}
	\end{equation}
	where $J=-2t^2/U<0$.  The first term in Eq. \eqref{Hamiltonian_T} is the $\SU(3)$ exchange interaction, and the second is the generalized DM interaction.
	
	Let us discuss the effect of the presence of the Zeeman term
	\begin{equation}
		\hH_{\rm Zeeman} = -h \sum_i \hS_i^z,
		\label{Hamiltonian_Zeeman}
	\end{equation}
	where $h$ is a coupling constant, and the spin-$1$ operator is given by 
	\begin{equation}
		\hS_i^a = \sum_{\sigma,\rho} \hb_{i,\sigma}^\dagger(\tau_a)_{\sigma\rho}\hb_{i,\rho},
	\end{equation}
	with the conventional spin-$1$ matrices $\tau_a$ $(a=x,y,z)$. 
	The fact that the operation of $\hS^z_i$ does not change the particle number implies $P_\calS [\hH_t^\pm, \hH_{\rm Zeeman}]P_\calS=0$. In addition, the term $[\heta,[\heta, \hH_{\rm Zeeman}]]\in\calO(t^2h/U^2)$ is negligible if $|h| \ll U$.
	Therefore, if we add the Zeeman term \eqref{Hamiltonian_Zeeman} into Eq. \eqref{Hamiltonian_gBH}, we just obtain the effective Hamiltonian \eqref{eq:H} given by Eq.~\eqref{Hamiltonian_T} with Eq.~\eqref{Hamiltonian_Zeeman} \citeSM{pixley2017strong}.

\bibliographystyleSM{apsrev4-1}
\bibliographySM{refs}

\end{document}